\documentclass[reprint,showkeys,floatfix,pra,
superscriptaddress,
amsmath,
amssymb,
aps,
nofootinbib
]{revtex4-2}

\usepackage{graphicx,setspace}% Include figure files
\usepackage{bm}% bold math
\makeatletter
\def\label#1{\@bsphack
  \begingroup
  \UseHookWithArguments{label}{1}{#1}%
  \protected@write\@auxout{}%
    {\string\newlabel{#1}{{\@currentlabel}{\thepage}%
    {\@currentlabelname}{\@currentHref}{\@kernel@reserved@label@data}}}%
  \endgroup
  \@esphack}
\makeatother
\usepackage{hyperref}% add hypertext capabilities
\usepackage{etoolbox}
\makeatletter
\patchcmd{\@bibdataout@aps}{author="08"}{author="48"}{}{}
\patchcmd{\@bibdataout@aps}{author="08"}{author="48"}{}{}
\makeatother
\begin{document}

\title{Evidence for Counterfactual Violation of Local Conservation Laws in Quantum Events}

%\author*[1,2]{\fnm{Patrick} \sur{Cameron} \email{patrick.cameron@unina.it}}
%\author[1]{\fnm{Francesco} \sur{Di Colandrea}}
%\author[1]{\fnm{Filippo} \sur{Cardano}}
%\author*[1,2]{\fnm{Lorenzo} \sur{Marrucci} \email{lorenzo.marrucci@unina.it}} 
%\affil[1]{\orgdiv{Department of Physics ``Ettore Pancini''}, \orgname{Università di Napoli Federico II}, \orgaddress{\city{Naples}, \country{Italy}}}
%\affil[2]{\orgname{Scuola Superiore Meridionale}, \orgaddress{\city{Naples}, \country{Italy}}}
\author{Patrick Cameron}
\email[Corresponding author: ]{p.cameron@ssmeridionale.it}
\affiliation{Department of Physics ``Ettore Pancini'', Universit{\`a} di Napoli Federico II, Naples, Italy}
\affiliation{Scuola Superiore Meridionale, Naples, Italy}

\author{Francesco Di Colandrea}
\affiliation{Department of Physics ``Ettore Pancini'', Universit{\`a} di Napoli Federico II, Naples, Italy}

\author{Filippo Cardano}
\affiliation{Department of Physics ``Ettore Pancini'', Universit{\`a} di Napoli Federico II, Naples, Italy}

\author{Lorenzo Marrucci}
\email[Corresponding author: ]{lorenzo.marrucci@unina.it}
\affiliation{Department of Physics ``Ettore Pancini'', Universit{\`a} di Napoli Federico II, Naples, Italy}
\affiliation{Scuola Superiore Meridionale, Naples, Italy}

%\begin{document}

%\abstract{Physical conservation laws, such as those of energy and momentum, are generally believed to hold exactly and locally in spacetime, including in quantum phenomena. Yet Aharonov, Popescu, and Rohrlich (APR) recently argued, on the basis of a thought experiment, that individual quantum events, as opposed to ensemble averages, may occasionally violate local conservation laws. Their argument relies on the wave phenomenon known as ``superoscillations'', which APR themselves discovered more than 30 years ago. Here we provide direct experimental evidence for such a local violation. We extract photons from a small superoscillatory region of a photonic field near the core of an optical vortex and show that their average transverse momentum is statistically incompatible with the predictions of local momentum conservation, even after taking into account the effect of the extraction mechanism. We also detect high-transverse-momentum photons at a rate significantly higher than that predicted by local conservation. Because this violation can be demonstrated only counterfactually and through postselection, it has no implications for relativistic spacetime causality. This result may represent the first example of a new class of quantum-nonlocal phenomena that do not explicitly rely on entanglement.}

\begin{abstract}
Physical conservation laws, such as those of energy and momentum, are generally believed to hold exactly and locally in spacetime, including in quantum phenomena. Yet Aharonov, Popescu, and Rohrlich (APR) recently argued, on the basis of a thought experiment, that individual quantum events, unlike ensemble averages, may occasionally violate local conservation laws. Their argument relies on the wave phenomenon known as ``superoscillations'', which APR themselves discovered more than 30 years ago. Here we provide experimental evidence for such a violation. We extract photons from a small superoscillatory region near the core of an optical vortex and show that their mean transverse momentum is statistically incompatible with a general bound implied by local momentum conservation. The derivation of this bound requires only the theoretically well-supported assumption that the extraction mechanism does not alter the photons’ mean transverse momentum. We also detect photons with high transverse momentum at a rate significantly exceeding that predicted by a model assuming local momentum conservation. Because this violation can be established only counterfactually and through postselection, it does not conflict with relativistic causality. Our results may represent the first example of a distinct form of quantum nonlocality that does not explicitly rely on entanglement.
\end{abstract}


\maketitle

%\clearpage

\section{Introduction}
Superoscillations occur when a signal or wave locally oscillates faster than the highest-frequency components in its Fourier spectrum \cite{berry_faster_1994,kempf_black_2000}. A prototypical example was proposed by Aharonov, Popescu, and Rohrlich (APR) in 1991 \cite{aharonov_soft_1991}: a wave with a finite frequency bandwidth, say between $-\omega_B$ and $\omega_B$, can oscillate within a certain spatiotemporal region approximately as a pure harmonic signal with frequency $\omega_S \gg \omega_B$. This superoscillatory region can, in principle, be made arbitrarily wide. The price to pay is that the superoscillating part of the signal is exponentially small compared with the surrounding Fourier-limited oscillations.

Superoscillations have attracted considerable attention in optics and other areas of physics and engineering, mainly because of their potential for superresolution imaging and other metrological applications \cite{berry_evolution_2006,ferreira_nyquist_2006,berry_roadmap_2019,chen_superoscillation_2019,zheludev_optical_2022,jordan_superoscillations_2025}. Several studies, in both the classical and quantum regimes, have already demonstrated the phenomenon experimentally (e.g., \cite{huang_optical_2007,rogers_lens_2012,yuan_quantum_2016,ghosh_azimuthal_2023,ma_observation_2026}). However, very few have focused on its implications for conservation laws. Yet a clear tension between superoscillations and conservation laws was already present in the original 1991 APR paper, titled ``How a soft photon can emit a hard photon'', which highlighted an apparent violation of energy conservation \cite{aharonov_soft_1991}. The existence of an actual violation has been debated in the literature (see, e.g., \cite{kempf_unusual_2004,berry_escaping_2018,afanasev_superkicks_2022}).

More recently, APR returned to these issues in two papers that examine the question of conservation laws in greater depth \cite{aharonov_conservation_2021,aharonov_conservation_2023}. They consider a particle confined in a cavity and initially prepared in a quantum state described by a superoscillatory wavefunction with energies $\le \hbar\omega_B$, where $\hbar$ is the reduced Planck constant. A suitable mechanism then extracts the particle selectively from the superoscillatory region of the cavity. Since the superoscillations of the wavefunction have very small amplitude, the probability of extraction is very low; in most repetitions of the experiment, the particle remains in the cavity and its wavefunction undergoes only a minimal perturbation. However, in the rare events in which the particle is extracted, the resulting superoscillatory wavefunction becomes very similar to that of a free particle with very large energy $\hbar\omega_S\gg\hbar\omega_B$. On average, one can prove that the mean energy of the particle, including both extraction and non-extraction events, is conserved, in agreement with standard theorems of quantum physics. In an individual quantum event in which the particle is extracted, however, the quantum state is reduced (collapsed) to the extracted-particle component, so that the non-extracted alternative no longer contributes; in more pictorial terms, the non-extracted part of the wavefunction ``disappears''. This leads to an apparent violation of energy conservation. An obvious objection is that, because the extraction mechanism is strongly localized in space and time, it must be associated with a broad distribution of energies that could be transferred to the particle during extraction, thereby reconciling the phenomenon with local energy conservation. On the other hand, by making the superoscillatory region sufficiently wide through a suitable choice of the particle wavefunction, this energy exchange can in principle be made arbitrarily small, although at the cost of further reducing the extraction probability. Moreover, using a detailed model of a specific extraction mechanism, APR show that the extraction process does involve a random, nonsystematic exchange of energy with the particle, but that this exchange is merely additional and bears no relation to the deterministic superoscillatory energy of the extracted particle.

\begin{figure*}[th]
    \centering
    \includegraphics[width=\textwidth]{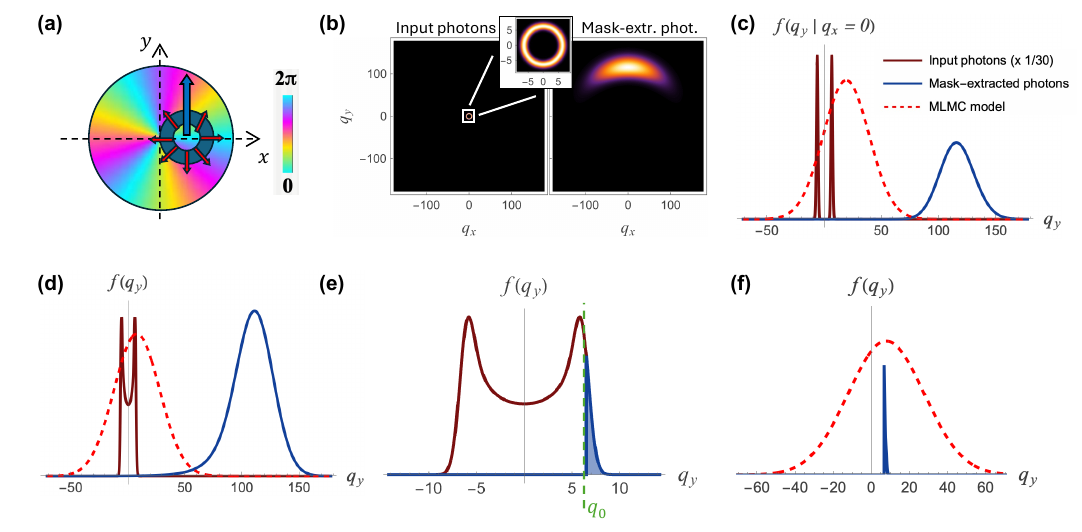}
    \caption{Concept of the momentum non-conservation experiment and an idealized case. {\bf(a)} A mask is used to extract photons from a small spatial region of the input vortex wavefunction, close to the vortex core. The extracted photons exhibit a large ``superkick'' momentum in the direction of the local phase gradient (large blue vertical arrow), superimposed on random momentum kicks due to the mask (smaller red arrows). All momentum scales are in units of $\hbar/w_0$. {\bf(b)-(d)} show the calculated photon behavior for an input vortex charge $\ell=20$ and a Gaussian extraction mask with $w_M = 0.05 w_0$ centered at $x_M=d_M=0.03 w_0$. {\bf(b)} Two-dimensional (2D) normalized momentum distributions for the input photons (left; the inset shows an enlarged view of the central region) and the mask-extracted photons (right), showing the large average $q_y$-momentum of the mask-extracted photons. {\bf(c)} Vertical cut of the same distributions at $q_x=0$. The dark red line shows the input LG distribution (divided by 30 for viewing clarity), whereas the blue line shows the distribution of mask-transmitted photons. The red dashed curve gives the MLMC distribution (see main text and panels {\bf(e)-(f)}). {\bf(d)} Corresponding $q_x$-integrated marginal distributions of the momentum $q_y$. In this case, detecting a single photon with, say, $q_y>120 \hbar/w_0$ would be enough to reject the MLMC null hypothesis with high statistical significance ($p\mathrm{-value}\simeq 2\times 10^{-7}$). Unfortunately, the mask transmission probability for this idealized example is $P_{\mathrm{tr}} = 5\times10^{-53}$, clearly unreachable experimentally. Panels {\bf(e)-(f)} illustrate the construction of the MLMC model: {\bf(e)} from the total input photon $q_y$ distribution (dark red line), the mask is assumed to select for transmission only photons with $q_y>q_0$ (blue line), where $q_0$ is determined by imposing that the total selected probability (light blue area under the line) is equal to the mask transmission one $P_{\mathrm{tr}}$. {\bf(f)} The transmitted photons then exchange an additional random momentum with the mask, with zero mean, leading to their final MLMC distribution (dashed red curve).}
    \label{fig:concept}
\end{figure*}

The same arguments can be readily extended to other conservation laws associated with spacetime symmetries, including those of momentum and angular momentum. In the second of these papers, APR present another example involving angular momentum and, more importantly, put forward a possible entanglement-based mechanism that could restore the validity of conservation laws in individual quantum events, although at the cost of accepting a form of (counterfactual) nonlocality \cite{aharonov_conservation_2023}. We return to this APR proposal in the Discussion section.

In the work reported here, we move from the thought experiments proposed by APR to a laboratory experiment aimed at demonstrating that a conservation law---specifically, momentum conservation along a given direction---can be locally violated in individual postselected quantum events. As we show below, our experiment is based on the superoscillatory behavior of light fields near the core of an optical vortex \cite{nye_dislocations_1974,berry_waves_2008}. In this region, optical vortices are associated with large transverse gradients of optical phase, corresponding to local spatial frequencies that can become arbitrarily large and hence exceed the transverse Fourier spectrum \cite{berry_five_2013}. The fact that optical vortices become superoscillatory near their cores was first pointed out by M.\ V.\ Berry \cite{berry_waves_2008}. Berry, together with S.\ M.\ Barnett, also suggested that the momentum transferred to atoms localized in this superoscillatory region when they absorb a photon could exceed any of the momenta carried by the vortex photons \cite{berry_five_2013,barnett_superweak_2013}: a ``superkick'', as they named it. However, the strong localization typically associated with atomic wavefunctions implies a very broad momentum distribution from the outset, making a violation of the conservation law difficult to demonstrate in that context.

Rather than attempting this challenging atomic experiment, we follow the same general idea but apply it directly to the photons, in close analogy with APR's original proposal to introduce an extraction mechanism for the superoscillatory particle. In our case, extraction is implemented by a small transmission mask positioned near the vortex core, as shown schematically in Fig.~\ref{fig:concept}{\bf(a)}, and we examine the transverse momentum distribution of the transmitted photons. Figure \ref{fig:concept}{\bf(b)--(d)} compares a calculated transmitted-photon distribution with both the corresponding input-photon distribution and a \textit{maximal local-momentum-conserving} (MLMC) reference distribution, where ``maximal'' means maximally biased toward high momentum values (see below for a precise definition). The extracted-photon distribution overlaps only marginally with the other two, indicating a clear violation of local momentum conservation. This is, however, a highly idealized case involving a high-order vortex and a very small mask positioned close to the vortex core. Such a regime is experimentally inaccessible because of the vanishingly small mask transmission probability.

An experimentally realizable case is instead obtained using a first-order vortex for the input photons, with the mask size and position optimized for signal-to-noise ratio. As we show below, our calculated and experimental results indicate that the mask-extracted photons still exhibit an average momentum along a given transverse direction that is significantly larger than the maximum allowed by local momentum conservation. In addition, we observe a significant excess of high-momentum photons. These two statistical tests provide evidence that local momentum conservation is violated within the postselected subset of quantum events corresponding to photons transmitted through the mask.

\section{Results}
\subsection{Experimental concept and modeling}
Consider a photon whose vortex wavefunction is described, in the paraxial approximation, by a monochromatic Laguerre-Gauss (LG) mode propagating along the $z$ axis, with radial index $p=0$ and vortex charge $\pm\ell$ (LG$_{0,\pm\ell}$ mode):
\begin{equation}
    \psi(\bm{r},t)
    = \frac{N_\ell}{w_0^{\ell+1}} (x \pm i y)^\ell
    e^{-\frac{x^2+y^2}{w_0^2}}.
\label{eq_inputwave}
\end{equation}
Here $w_0$ is the mode waist and $N_\ell = \sqrt{2^{\ell+1}/(\pi \ell!)}$ is a normalization constant. For brevity, here and in the following we omit the propagation $z$ and time dependences and refer only to the transverse $xy$ wavefunction. We also omit the uniform fixed polarization state, which plays no role in our experiment. The vortex charge $\ell$ is associated with the orbital angular momentum (OAM) of the photon, given by $\pm\ell\hbar$. The $\pm$ sign distinguishes positive and negative vortex phase circulations and OAM values.

\begin{figure}[tbh]
    \centering
    \includegraphics[width=0.8\linewidth]{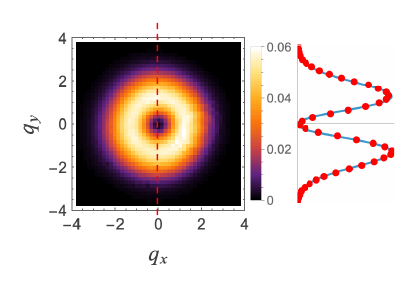}
    \caption{Input LG mode used for our experiments, with vortex charge $\ell=1$. Experimental normalized two-dimensional momentum distribution (that is, spatial distribution in the focal plane of an imaging lens) obtained in the photon-counting regime. Momentum units are $\hbar/w_0$. Brightness is proportional to the photon counts per pixel. The side panel shows the vertical cut through the vortex for $q_x=0$: the red dots are data (the error bars are smaller than the symbols) and the blue curve is theory. No fitting parameters are used.}
    \label{fig:inputLG}
\end{figure}

The momentum-space wavefunction of the input photon is given by
\begin{equation}
    \tilde{\psi}(\bm{q})
    = (-i)^\ell \frac{N_\ell}{w_q^{\ell+1}}
    (q_x \pm i q_y)^\ell
    e^{-\frac{q_x^2+q_y^2}{w_q^2}},
\label{eq_inputmomentumwave}
\end{equation}
where $\bm{q}=(q_x,q_y)$ is the transverse momentum vector and $w_q=2\hbar/w_0$ is the width of the momentum distribution. The normalized momentum distribution is then $f(\bm{q})=|\tilde{\psi}(\bm{q})|^2$ and is invariant under free-space propagation along $z$. The mean input momentum $\bar{\bm{q}}_{\mathrm{in}}$ corresponding to Eq.~(\ref{eq_inputmomentumwave}) vanishes by symmetry. An example of a measured LG momentum distribution for $\ell=1$ is shown in Fig.~\ref{fig:inputLG}. Among all possible vortex-mode profiles, we choose an LG mode because it has the narrowest possible momentum distribution for a given spatial extension. To prove strict superoscillatory behavior, one might ideally use a bounded-support momentum distribution for the vortex photons, for example by low-pass filtering the input beam in a suitable Fourier plane. However, such a filtered beam would exhibit large power-law radial tails that would be truncated by any subsequent finite-aperture optics in the setup, thereby reintroducing an unbounded distribution of transverse momentum and making the whole procedure ineffective. We therefore adopt a generalized definition of superoscillations that does not require a bounded-support Fourier spectrum \cite{berry_leaky_2019}.

\begin{figure*}[t]
    \centering
    \includegraphics[width=\textwidth]{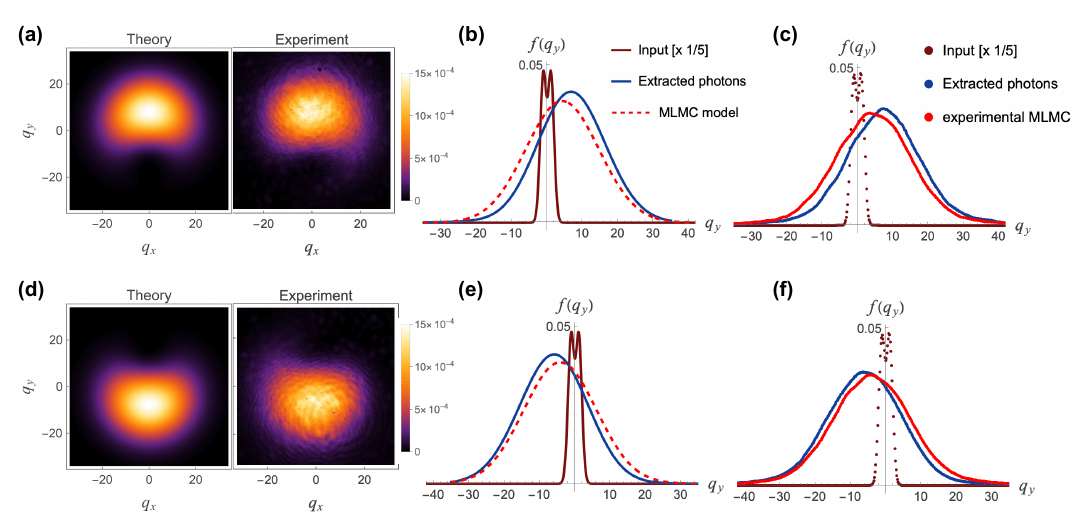}
    \caption{Main experimental results, providing evidence for a violation of local momentum conservation. The input vortex has charge $\ell=1$, and the extraction mask has nominal values $w_M=0.1w_0$ and $d_M=0.1w_0$, corresponding to a mask transmission probability $P_{\mathrm{tr}}=2.7\times10^{-4}$. All momentum scales are in units of $\hbar/w_0$. {\bf(a)} Theoretical (left) and measured (right) 2D normalized momentum distributions of photons transmitted by the mask, showing the ``superkick'' average momentum $\bar{q}_{Ty}>0$. {\bf(b)} Computed marginal distributions of $q_y$ for the input photons (brown solid line, divided by 5 for visibility), the MLMC null hypothesis (red dashed line), and the mask-extracted photons (blue solid line). No adjustable parameters are used in this calculation. {\bf(c)} Measured marginal distributions of $q_y$ for the input photons (brown dots, divided by 5), the experimental reconstruction of the MLMC model (red dots), and the mask-extracted photons (blue dots). Error bars are smaller than the symbols. Panels {\bf(d)}-{\bf(f)} show the corresponding results for a mask positioned on the opposite side of the vortex at nominal position $x_M=-d_M=-0.1w_0$. The measured mask properties are: $w_M=(0.097\pm0.001)w_0$; $x_M=d_M=(0.10\pm0.01)w_0$ in {\bf(c)}, $x_M=-d_M=(-0.13\pm0.01)w_0$ in {\bf(f)}. The calculations in panels {\bf(b)} and {\bf(e)} use these measured mask parameters.}
    \label{fig:mainresults}
\end{figure*}

We now place a transmission mask along the path of the photon. We choose a Gaussian transmission mask of width $w_M$, centered a distance $d_M$ from the vortex core at $x=y=0$. The Gaussian profile is again chosen to minimize the width of the associated momentum distribution. Without loss of generality, we place the mask center on the positive $x$ axis, i.e. $x_M=d_M,y_M=0$ (we later consider also $x_M=-d_M$). The photon wavefunction in real space is therefore multiplied by the following mask amplitude-transmission function: $M(\bm{r}) = e^{-[(x-d_M)^2+y^2]/w_M^2}$. The mask-transmitted (or extracted) photons are then described by a new momentum-space wavefunction $\tilde{\psi}_T(\bm{q})$, which can be computed analytically (it is essentially a scalar-wave Fraunhofer diffraction calculation; see Appendix \ref{app:main-modeling} for details). The associated momentum distribution $f_T(\bm{q}) = |\tilde{\psi}_T(\bm{q})|^2$ is
\begin{equation}
\begin{aligned}
    f_T(\bm{q}) & = \frac{
        N_\ell^2 w_q^{2\ell+2}
        e^{
            -\frac{d_M^2 w_q^2 w_{Mq}^2}
            {2\hbar^2\left(w_q^2+w_{Mq}^2\right)}
        }
    }{
        \left(w_q^2+w_{Mq}^2\right)^{2\ell+2}
    } \\
    &\times \left[
        q_x^2+
        \left(
            \frac{d_M w_{Mq}^2}{2\hbar}
            \pm q_y
        \right)^2
    \right]^\ell
    e^{
        -\frac{2\left(q_x^2+q_y^2\right)}
        {w_q^2+w_{Mq}^2}
    },
\label{eq_maskedmomentumdistr}
\end{aligned}
\end{equation}
where $w_{Mq}=2\hbar/w_M$. This $f_T$ is unnormalized because it includes the probability of mask transmission; to normalize, it must be divided by the transmission probability $P_{\mathrm{tr}}=\iint_{\mathbb{R}^2} |\psi_T(x,y)|^2\,dx\,dy$ (see Appendix \ref{app:main-modeling}). Examples of calculated input and mask-transmitted momentum distributions are shown in Figs.\ \ref{fig:concept}{\bf(b)-(d)} and \ref{fig:mainresults}{\bf(b)}.

From Eq.~(\ref{eq_maskedmomentumdistr}), one can compute the average $y$-momentum of the photons transmitted by the mask. For $\ell=1$ one finds:
\begin{equation}
\bar{q}_{Ty} = \frac{\pm \hbar/d_M}
{\frac{w_0^2}{w_0^2+w_M^2}+\frac{w_M^2}{2d_M^2}}.
\label{eq_meanmomentum}
\end{equation}
This is the ``superkick'' momentum acquired by the photons extracted from the superoscillatory region. In principle, it can be made arbitrarily large in the joint limit $w_M \ll w_0$ and $d_M \ll w_0$ (within the validity of paraxial approximation). If one imposes the additional condition $w_M \ll d_M$ (not required for our purposes), one obtains $\bar{q}_{Ty}\simeq \pm\hbar/d_M$, which coincides with the local momentum as defined in five alternative ways by Berry \cite{berry_five_2013}, including the ``weak value'' of the photon momentum when postselected for the position of the mask center.

We now introduce \textit{assumption A1}: this deterministic contribution to the mean momentum is \textit{not} supplied by the mask (the extraction device). This assumption is supported theoretically by a simple quantum model of the photon--mask interaction, which predicts a vanishing mean recoil of the mask (see Appendix \ref{app:supp-mask-recoil}). Alternatively, one can use the following qualitative argument, which is independent of any specific model: if the transmission mask were replaced by an absorbing one, the same superoscillatory momentum $\bar{q}_{Ty}$ would be transferred from the absorbed photon to the mask, in analogy with atomic superkicks. In this case, however, the mask would be the only relevant final system, and there would be no other local bodies that could compensate this nonzero average momentum.

For momentum conservation to hold locally, under assumption A1, momentum values around $\bar{q}_{Ty}$ should already be present in the input photon distribution, and the mask should somehow preferentially transmit photons with such momenta. This selection would bias the momentum distribution of the transmitted photons and possibly give rise to the observed nonzero average momentum. Let us now determine the maximum mean momentum that such a selective-mask theory can predict for the extracted photons, subject to a fixed mask transmission probability $P_{\mathrm{tr}}$. This maximum is equal to the average $\bar{q}_{Sy}$ obtained from the upper- or lower-tail distribution $f_S(\bm{q})=N_S f(\bm{q})\theta(\pm q_y-q_0)$, where $\theta$ denotes the Heaviside step function, $N_S$ is a normalization constant, and $q_0$ is a momentum threshold chosen so that
\begin{equation}
\iint f(\bm{q})\theta(\pm q_y-q_0)\,d^2q=P_{\mathrm{tr}}
\end{equation}
(see Fig.~\ref{fig:concept}{\bf(e)}). The mask may further add a random kick to the momentum-selected photons, but under A1 this kick cannot change the mean momentum of the transmitted photons. We conclude that a local-momentum-conserving theory, under the sole assumption A1, must generally satisfy the inequality $|\bar{q}_{Ty}|\leq |\bar{q}_{Sy}|$. A statistically-significant violation of this inequality, namely an average superkick momentum larger than $|\bar{q}_{Sy}|$, would therefore constitute evidence for a violation of local momentum conservation (conditional to assumption A1).

Besides the average, we can consider the full momentum distribution predicted by a local-momentum-conserving distribution that is maximally biased toward large $|q_y|$ in the direction of the superkick (MLMC). Following a similar reasoning as that above, this distribution is $f_{\mathrm{MLMC}}(\bm{q})=K_M(\bm{q}) \ast f_S(\bm{q})$, namely the convolution of the renormalized upper- or lower-tail distribution $f_S(\bm{q})$ with the mask-random-kick distribution $K_M(\bm{q})$ (see Fig.~\ref{fig:concept}{\bf(f)}). Here we introduce \textit{assumption A2}: the distribution of random kicks induced by the mask is independent of the selected photon's input momentum. This is consistent with the known behavior of a mask in the paraxial limit when a single-momentum plane wave is used as input. Besides the bound $|\bar{q}_{Sy}|$ on the average momentum, this MLMC distribution will be our main reference distribution, or null hypothesis, for testing the local violation of momentum conservation. For a suitable choice of mask parameters, we find that the upper or lower tail of the transmitted-photon distribution $f_T(\bm{q})/P_{\mathrm{tr}}$ in the direction of the superkick contains values of $q_y$ that are significantly more probable than expected from the MLMC distribution. This excess of high-momentum photons can therefore provide the basis for a further experimental test of local momentum conservation (conditional to both assumptions A1 and A2).

\subsection{Experimental results}
Our main experiment is performed in the photon-counting regime, which allows measuring low single-photon probabilities with high precision. Experimental details are given in Methods and Appendix \ref{app:supp-setup}. Figure \ref{fig:mainresults} shows two representative experimental datasets and compares them with theory, showing very good qualitative agreement. We emphasize that there are no adjustable parameters in these theoretical curves. Both data and theory are normalized and the momentum origin and momentum-per-pixel scales are both fixed by the observed input LG mode momentum distribution (additional details are given in Methods and Appendix \ref{app:supp-data-analysis}). The MLMC reference distribution was also experimentally reconstructed using the procedure described in Methods. In doing so, we introduce \textit{assumption A3}: the very weak ($\sim10^{-4}$) power-law pedestal that unavoidably accompanies the input LG momentum distribution and is attributed to finite-aperture optical apodization is \textit{not} preferentially selected by the mask when constructing $f_S(q_y)$ (see Methods and Appendix \ref{app:supp-data-analysis}). This is physically justified, in our view, because such apodization occurs far from the mask location and has no relationship with the vortex phase gradients that would give rise to the preferential momentum selection. 

\begin{table*}[tbh]
    \centering
    \caption{Average momentum $\bar{q}_{Ty}$ and number of high-momentum photons $N_{\mathrm{ph}}$ for mask-transmitted photons, compared with the local-conservation bound $\bar{q}_{Sy}$ and with the experimental MLMC model, for the same data shown in Fig.~\protect\ref{fig:mainresults}{\bf(c), (f)}. All momentum values are in units of $\hbar/w_0$. High-momentum is here defined by being above a threshold-momentum of $\pm5\hbar/w_0$ or $\pm15\hbar/w_0$ (in the same direction as the superkick). The first three rows of the table refer to $x_M=d_M=(0.10\pm0.01)w_0$; the total number of detected photons was $1.38\times10^8$. The last three rows refer to $x_M=-d_M=(-0.13\pm0.01)w_0$; the total number of detected photons was $2.07\times10^8$. All uncertainties represent one standard deviation.}
    \label{tab:statistics}
    %\small
    %\footnotesize
    \setlength{\tabcolsep}{12pt}
    \begin{tabular}{ccccc}
        \hline
        Statistic & Theory & Experiment & $\bar{q}_{Sy}$ / MLMC bounds & Exp.$-$ $\bar{q}_{Sy}$/MLMC \\
        \hline
        $\bar{q}_{Ty}$ & $6.84$ & $7.2\pm0.3$ & $4.2\pm0.1$ & $3.0\pm0.3$ \\
        $N_{\mathrm{ph}}(q_y > 5)$ & $7.96\times10^7$ & $(8.17\pm0.02)\times10^7$ & $(6.68\pm0.01)\times10^7$ & $(1.49\pm0.02)\times10^7$\\
        $N_{\mathrm{ph}}(q_y > 15)$ & $2.74\times10^7$ & $(3.24\pm0.02)\times10^7$ & $(2.35\pm0.01)\times10^7$ & $(0.90\pm0.02)\times10^7$\\
        \hline
        $\bar{q}_{Ty}$ & $-6.06$ & $-6.0\pm0.3$ & $-4.3\pm0.1$ & $-1.7\pm0.3$ \\
        $N_{\mathrm{ph}}(q_y <-5)$ & $11.19\times10^7$ & $(10.82\pm0.03)\times10^7$ & $(9.50\pm0.02)\times10^7$ & $(1.32\pm0.03)\times10^7$\\
        $N_{\mathrm{ph}}(q_y <-15)$ & $3.68\times10^7$ & $(4.12\pm0.03)\times10^7$ & $(3.36\pm0.02)\times10^7$ & $(0.76\pm0.04)\times10^7$\\
        \hline
    \end{tabular}
\end{table*}

Table \ref{tab:statistics} reports the test statistics used to prove the violation of the average-momentum bound $|\bar{q}_{Ty}|\leq |\bar{q}_{Sy}|$ and to reject the MLMC-distribution null hypothesis. In both datasets, the measured average momentum lies beyond the inequality bound in the predicted superkick direction by more than five standard deviations (see Methods for uncertainty estimation); the larger of the two corresponding two-sided Gaussian $p$-values is $1.5\times10^{-8}$. The observed numbers of high-momentum photons (for two different definitions of ``high'') exceed the MLMC predictions by several tens of standard deviations, corresponding to negligibly small Gaussian $p$-values.

\begin{figure}[!b]
    \centering
    \includegraphics[width=0.8\linewidth]{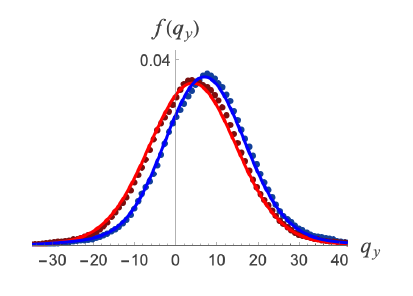}
    \caption{Comparison of the experimental momentum distributions shown in Fig.~\ref{fig:mainresults}{\bf(c)} with a model that also includes a small broadening effect attributed to optical apodization by finite-diameter imaging optics in intermediate propagation planes downstream of the mask. The data are downsampled for visual clarity. The two parameters of this model (see Appendix \ref{app:supp-data-analysis}) are obtained by fitting the data. The blue dots and line represent the mask-transmitted photons; the red dots and line represent the MLMC model. Error bars are smaller than the symbols.}
    \label{fig:theory_data_match}
\end{figure}

Accounting for optical apodization introduced by the imaging optics explains the small quantitative discrepancies between theory and experiment seen in Fig.~\ref{fig:mainresults}. As shown in Fig.~\ref{fig:theory_data_match}, a model including two adjustable parameters that characterize this apodization yields excellent quantitative agreement (see Appendix \ref{app:supp-data-analysis} for details).

\section{Discussion}
Our experimental results provide evidence for a local violation of momentum conservation for individual photons extracted from the superoscillatory region of an optical vortex, under a few physically justified assumptions (A1 and A3 for the violation of the mean-momentum inequality; A1, A2, and A3 for the excess of high-momentum photons). This verification is, however, intrinsically counterfactual: for the same photon, we cannot measure both the input momentum distribution and the momentum distribution after transmission through the mask. We therefore compare momentum distributions obtained in two mutually exclusive experimental configurations, in a way that is conceptually analogous to the use of alternative measurement settings in tests of Bell inequalities. The inference of a violation relies on taking the input momentum distribution predicted by the prepared vortex wavefunction as the relevant counterfactual distribution for the runs in which the mask-transmission measurement is performed. We note, however, that this assumption is weaker than the local-realist assumptions used in Bell tests: the derivation of the local conservation bounds does not require assigning definite pre-existing momenta to individual photons, but only the validity of the ensemble momentum distribution associated with the prepared input state.

Because transverse momentum distributions are preserved under free-space propagation, the relevant momentum measurements could, in principle, be performed far from the preparation plane where the superoscillatory vortex state is generated (in the present experiment, this distance was approximately one meter). The measurement configuration could therefore be selected only after the photons have left the source, in a spacetime arrangement such that no subluminal signal carrying that choice could return to the source in time to affect the prepared momentum distribution. Such a delayed-choice arrangement could be tested in future experiments.

One could in principle choose parameters for which the detection of a single mask-transmitted photon with high momentum would be sufficient to reject the MLMC null hypothesis with high statistical significance (this is the case of the example shown in Fig.~\ref{fig:concept}{\bf(b)-(d)}), although such a regime is experimentally out of reach. Our experiment instead involves a large number of photons: the violation of momentum conservation identified here is therefore a conditional statement about the subset of individual postselected quantum events in which the photon is actually transmitted through the mask, while there is no violation if photons blocked by the mask are included, or for a classical description of the whole optical field. The conditional character of the violation does not make it merely formal, however. In a single experiment involving a finite number $N$ of input photons, the number $N_T$ of transmitted photons fluctuates randomly, so the momentum compensation with the $N-N_T$ blocked photons would be imperfect and lead to a fluctuation of the total momentum of the system. This fluctuation could in principle be observable. As an extreme illustration, although with exceedingly small probability, all $N$ photons could be transmitted by the mask, in which case the momentum unbalance would apply to the entire $N$-photon output of the experiment.

The apparent local violation of conservation laws leaves two broad interpretive possibilities: either conservation laws are not strictly enforceable at the level of individual quantum events and hence are valid only on average, or they are enforced through a nonlocal mechanism involving additional physical systems. APR have proposed a mechanism of the latter kind, which could restore exact conservation in individual events \cite{aharonov_conservation_2023} (see also \cite{collins_conservation_2025,collins_networks_2026}). In their picture, preparing a particle in a specified wavefunction necessarily establishes correlations---more precisely, a small entanglement in the relevant position and momentum/energy degrees of freedom---between the particle and the preparation apparatus, or more generally the (quantum) reference system that defines the state. In our optical implementation, analogous correlations would involve the input photons and the spatial light modulator (SLM) used to define the vortex mode. APR then argue that, although such correlations are negligible in ordinary situations, they cannot be ignored for superoscillatory wavefunctions: when a photon is extracted from the superoscillatory region, the measured superkick momentum is balanced by an opposite momentum change of the preparation or reference system. This momentum change is extremely small compared with the width of the reference system's momentum distribution, which must itself be large enough to allow the superoscillatory wavefunction to be defined precisely in the spatial domain.

Our experiment does not directly test this APR mechanism, because it does not monitor the compensating momentum change of the preparation or reference system. In the present implementation such a system (the SLM) is macroscopic, so any such momentum shift would be far too small for experimental accessibility. A more direct test would require a genuinely quantum reference system whose momentum can be measured. One possible route would be an experiment with two quantum systems---for example, two entangled photons---prepared in a suitable joint superoscillatory state, in which the superkick acquired by one subsystem could be correlated with a compensating momentum change of the other. Such an experiment could test whether the apparent violation observed here is balanced by a nonlocal accounting of total momentum, or more generally of conserved quantities, in individual quantum events.

\section{Methods}
\subsection*{Experimental apparatus}
Our photon source is a superluminescent light-emitting diode (SLED), with central wavelength of 810 nm, coupled to a single-mode fiber. A 3-nm bandwidth filter is used to define the wavelength more precisely. After the light is coupled out of the fiber, a first SLM followed by an imaging system is used to prepare an input LG beam with vortex charge $\pm 1$, following the method introduced by Bolduc et al.\ \cite{bolduc_exact_2013}. We then image the beam onto a second SLM. The input LG waist radius $w_0$ at this stage is set to a value between $1.50$ and $1.80$ mm. The second SLM can either implement the transmission mask or direct the whole beam toward the detection line. The mask is realized as a contrast-modulated binary diffraction grating. Its contrast amplitude has a Gaussian radial profile, with radius $w_M\simeq0.1w_0$, centered at $x_M=\pm d_M$ with $d_M\simeq0.1w_0$ relative to the input beam center. The photon-detection camera is placed in the Fourier plane of an imaging system, so as to reveal the transverse momentum distribution of the detected photons. For our main measurements, we operate in the photon-counting regime using a $512\times512$ single-photon avalanche diode (SPAD) array (PiImaging SPAD512). We attenuate the source so that photon detections are sparse and multi-photon events per pixel are negligible. More precise values of the mask parameters $w_M$ and $d_M$ are independently measured from the diffraction of a Gaussian input beam and from the modulations in the measured mask transmission probability, respectively. We often observe significant ($\sim 0.01-0.03 w_0$) deviations from the nominal value of $d_M$, which we ascribe to vortex displacements resulting from residual coherent background light coming from the first SLM (e.g., a $10^{-4}$ power fraction of a Gaussian residual is enough to induce a $d_M$ deviation of about $0.01w_0$).

\subsection*{Experimental MLMC distribution}
The experimental MLMC distribution is reconstructed in three steps. First, we determine the $q_y$-marginal mask momentum distribution $K_M(q_y)$ by measuring the momentum distribution transmitted through the mask for a Gaussian input beam. Second, from the measured input distribution $f(q_y)$, we retain either the upper tail, $q_y>q_0$, or the lower tail, $q_y<-q_0$, according to the direction of the superkick. The threshold $q_0$ is chosen so that the retained tail probability equals $P_{\mathrm{tr}}$. After renormalization, we obtain the selected experimental marginal distribution $f_S(q_y)$. Third, we numerically convolve $f_S(q_y)$ with $K_M(q_y)$. This procedure incorporates small deviations of the measured LG-mode tails from the ideal profile, which generally broaden the tails and therefore make the bound-violation verification more conservative.

As mentioned in the main text, we exclude from the construction of $f_S(q_y)$ a weak power-law pedestal observed in all our LG measurements. Its amplitude is less than $10^{-4}$ of the main signal per pixel and approximately $10^{-3}$ in the $q_y$ marginals. We attribute it to unavoidable finite-aperture apodization in intermediate propagation planes, possibly combined with interference from a weak coherent background. Over the measured range, its marginal distribution is well described by a dependence proportional to $1/|q_y|$, whose extrapolation is formally nonintegrable. Including this pedestal in $f_S(q_y)$ would therefore make its normalization and statistical properties depend on the arbitrary experimental window. Making the pedestal smaller would not solve the issue; at most, it could hide it below the background shot noise. Accordingly, both the mean-momentum inequality bound and the MLMC model require the assumption A3 that photons contributing to the pedestal tails are not preferentially transmitted by the mask. The full unselected measured pedestal is nevertheless retained, by adding it to $f_S(q_y)$, before calculating the final convolution used to construct the experimental MLMC distribution.

\subsection*{Uncertainty estimation}
Uncertainties based on Poisson counting statistics are negligible. The dominant uncertainties are instead estimated from the residual differences between the data and our full model, which includes optical apodization. Specifically, we assign a constant uncertainty to each point of the $q_y$-marginal distributions and choose its value so that the reduced $\chi^2$ is unity. This procedure empirically accounts for uncontrolled optical imperfections not captured by the model and provides an estimate of the residual experimental uncertainty. We also use Monte Carlo simulations to estimate the additional uncertainty introduced by the pedestal-subtraction procedure described above (see Appendix \ref{app:supp-data-analysis} for details). The measured mask-transmission probabilities are consistently slightly higher than the theoretical values, presumably because of residual background light. Because using the measured values would significantly lower the MLMC bounds, we conservatively use the theoretical values.

%\backmatter (poi cambia subsection in section nei tre paragrafi di backmatter)
%\vspace{2\baselineskip}

\section*{Acknowledgements}
We acknowledge financial support from the European Union via the Italian Ministry of University and Research (MUR), through the EU-funded PNRR MUR project PE0000023-NQSTI (National Quantum Science and Technology Institute).
\section*{Author contributions}
LM conceived the experiment and developed the associated modeling. LM, FC, and PC designed the experimental layout. PC assembled the setup and carried out the measurements. PC, FDC, and LM analyzed the data. FC supervised the experimental work. All authors discussed the results. LM wrote the manuscript, with feedback and contributions from all authors.
\section*{Competing interests}
The authors declare no competing interests.

\bibliography{references}

\clearpage
\onecolumngrid
\appendix
\numberwithin{equation}{section}

\section{Main theoretical modeling}
\label{app:main-modeling}

\subsection{Notations and description of the system}
\label{app:supp-notation}
Let us assume that the photon initial wavefunction is described by a single OAM eigenstate $\left\lvert \pm \left.\ell \right\rangle\right.$ with OAM eigenvalue $\pm \ell \hbar$, where $\ell$ is a non-negative integer, and in particular let us take a Laguerre-Gauss (LG) mode (in paraxial approximation) with radial index $p=0$:
\begin{equation}
\psi \left(\bm{r},t\right)=\psi \left(x,y,z,t\right)=\left\langle x,y,z\right.\left\lvert \pm \left.\ell \right\rangle\right.=\frac{{N}_{\ell }}{{w}_{0}^{\ell +1}}{\left(x\pm iy\right)}^{\ell }{e}^{-\frac{{x}^{2}+{y}^{2}}{{w}_{0}^{2}}}{e}^{i\left(kz-\omega t\right)},
\label{eq:supp-0-1}
\end{equation}
where ${N}_{\ell }$ is a dimensionless normalization constant, $k=\frac{2\pi }{\lambda }$ the wavenumber, with $\lambda$ the wavelength, ${w}_{0}$ is the beam waist, and we are assuming for simplicity that the Rayleigh length ${z}_{0}=\frac{\pi {w}_{0}^{2}}{\lambda }$ is very large so that the space dependence is well approximated by that valid around the focal region (that is, we neglect diffraction). The normalization constant (for our $p=0$ case) is the following:
\begin{equation}
{N}_{\ell }=\sqrt{\frac{{2}^{\ell +1}}{\pi \left(\ell !\right)}}.
\end{equation}
From now on, we will omit the $z$ and $t$ variables for brevity. In cylindrical coordinates, the same wavefunction is
\begin{equation}
\psi \left(r,\phi \right)=\left\langle r,\phi \right.\left\lvert \pm \left.\ell \right\rangle\right.=\frac{{N}_{\ell }}{{w}_{0}^{\ell +1}}{r}^{\ell }{e}^{\pm i\ell \phi }{e}^{-\frac{{r}^{2}}{{w}_{0}^{2}}}.
\label{eq:supp-0-2}
\end{equation}
The corresponding transverse momentum space OAM wavefunction (2D spatial Fourier transform of the real space wavefunction, using the pre-factors $\frac{1}{\sqrt{2\pi \hbar }}$ for each 1-dimensional Fourier Transform) can be written as follows:
\begin{equation}
\widetilde{\psi }\left(\bm{q}\right)=\left\langle {q}_{x},{q}_{y}\right.\left\lvert \pm \left.\ell \right\rangle\right.={\left(-i\right)}^{\ell }\frac{{N}_{\ell }}{{w}_{q}^{\ell +1}}{\left({q}_{x}\pm i{q}_{y}\right)}^{\ell }{e}^{-\frac{{q}_{x}^{2}+{q}_{y}^{2}}{{w}_{q}^{2}}}.
\label{eq:supp-0-3}
\end{equation}
In polar coordinates this is also
\begin{equation}
\widetilde{\psi }\left(\bm{q}\right)=\left\langle q,{\phi }_{q}\right.\left\lvert \pm \left.\ell \right\rangle\right.={\left(-i\right)}^{\ell }\frac{{N}_{\ell }}{{w}_{q}^{\ell +1}}{q}^{\ell }{e}^{\pm i\ell {\phi }_{q}}{e}^{-\frac{{q}^{2}}{{w}_{q}^{2}}}.
\label{eq:supp-0-4}
\end{equation}
The momentum probability distribution is given by the absolute square of these expressions, that is:
\begin{equation}
f\left(\bm{q}\right)={\left\lvert \widetilde{\psi }\left(\bm{q}\right)\right\rvert}^{2}=\frac{{N}_{\ell }^{2}}{{w}_{q}^{2\ell +2}}{\left({q}_{x}^{2}+{q}_{y}^{2}\right)}^{\ell }{e}^{-2\frac{{q}_{x}^{2}+{q}_{y}^{2}}{{w}_{q}^{2}}}.
\label{eq:supp-0-5}
\end{equation}
In these expressions we introduced the transverse momentum vector $\bm{q}=({q}_{x},{q}_{y})$, its polar coordinates $q,{\phi }_{q}$, and the transverse momentum distribution waist width ${w}_{q}=2\hbar /{w}_{0}$. However, it should be noted that the actual width of the momentum distribution scales also with $\ell$, as $\sim {w}_{q}\sqrt{1+\ell }$, just as the spatial distribution width scales as $\sim {w}_{0}\sqrt{1+\ell }$.

Let us now apply a transmission mask to the photon. We choose a Gaussian mask centered at distance ${d}_{M}$ from the \textit{z}-axis and of width ${w}_{M}$. For definiteness, we place the mask center on the positive \textit{x} semiaxis, so that the mask center has coordinates ${x}_{M}={d}_{M}, {y}_{M}=0$. Hence, the photon wavefunction in real space is multiplied by the following mask transmission-amplitude function:
\begin{equation}
M\left(\bm{r}\right)={e}^{-\frac{{\left(x-{d}_{M}\right)}^{2}+{y}^{2}}{{w}_{M}^{2}}}={e}^{-\frac{{r}^{2}+{d}_{M}^{2}-2{d}_{M}r\cos\phi }{{w}_{M}^{2}}}.
\label{eq:supp-0-6}
\end{equation}
The resulting (unnormalized) masked wavefunction is hence $\psi_T=M(\bm{r})\psi (\bm{r})$. We omit the full explicit spatial expression here as we will not be using it. Notice that \eqref{eq:supp-0-6} gives the mask amplitude transmission function, not the intensity, which is given by its square.

An important quantity that is determined by \eqref{eq:supp-0-6} and by the structure of the input beam is the \textbf{total mask transmission probability} for the photons of the input beam, which is given by
\begin{equation}
\begin{aligned}
P_{\mathrm{tr}}
&=\iint_{-\infty }^{+\infty } {\left\lvert \psi_T(x,y)\right\rvert}^{2}\,dx\,dy = \iint_{-\infty }^{+\infty } {\left\lvert M(\bm r)\psi(\bm r)\right\rvert}^{2}\,dx\,dy \\
&=\frac{w_M^{2+2\ell}}{\left(w_0^2+w_M^2\right)^{\ell+1}}
L_{-\ell-1}\!\left[
\frac{2d_M^2w_0^2}{w_M^2\left(w_0^2+w_M^2\right)}
\right]e^{-2d_M^2/w_M^2},
\end{aligned}
\label{eq:supp-0-7a}
\end{equation}
where ${L}_{n}$ are the Laguerre polynomials. For $\ell =1$ we obtain (after also replacing for ${N}_{\ell }$):
\begin{equation}
P_{\mathrm{tr}}=\frac{{w}_{M}^{2}\left(2{d}_{M}^{2}{w}_{0}^{2}+{w}_{0}^{2}{w}_{M}^{2}+{w}_{M}^{4}\right)}{{\left({w}_{0}^{2}+{w}_{M}^{2}\right)}^{3}}{e}^{-\frac{2{d}_{M}^{2}}{{w}_{0}^{2}+{w}_{M}^{2}}}.
\label{eq:supp-0-7b}
\end{equation}
It will be useful to look at the mask effect directly in momentum space. For this purpose, we need the Fourier-transform of the mask function (still using the pre-factors $\frac{1}{\sqrt{2\pi \hbar }}$ for each coordinate), which is as follows:
\begin{equation}
\widetilde{M}\left(\bm{q}\right)=\frac{2\hbar }{{w}_{Mq}^{2}}{e}^{-\frac{{q}_{x}^{2}+{q}_{y}^{2}}{{w}_{Mq}^{2}}}{e}^{-i\frac{{q}_{x}{d}_{M}}{\hbar }}=\frac{2\hbar }{{w}_{Mq}^{2}}{e}^{-\frac{{q}^{2}}{{w}_{Mq}^{2}}}{e}^{-i\frac{q{d}_{M}\cos{\phi }_{q}}{\hbar }},
\label{eq:supp-0-8}
\end{equation}
where we introduced the mask momentum distribution width ${w}_{Mq}=2\hbar /{w}_{M}$. If a particle with a well-defined initial transverse momentum (say in a very wide Gaussian wavefunction of waist ${w}_{L}$) passes through this mask, the unnormalized momentum distribution of the particle after the mask will be given by ${\frac{2}{\pi {w}_{L}^{2}}\left\lvert \widetilde{M}\left(\bm{q}\right)\right\rvert}^{2}$. If we instead normalize the wavefunction after the transmission through the mask (to obtain the distribution probability conditioned on the particle having passed the mask), then the resulting momentum distribution is given by
\begin{equation}
K_M\left(\bm{q}\right)\equiv{\left\lvert {\widetilde{M}}_{N}\left(\bm{q}\right)\right\rvert}^{2}=\frac{{w}_{Mq}^{2}}{2\pi {\hbar }^{2}}{\left\lvert \widetilde{M}\left(\bm{q}\right)\right\rvert}^{2}=\frac{2}{\pi {w}_{Mq}^{2}}{e}^{-2\frac{{q}_{x}^{2}+{q}_{y}^{2}}{{w}_{Mq}^{2}}},
\label{eq:supp-0-9}
\end{equation}
which also corresponds to the normalized mask-kick distribution $K_M$ used in the main text.

\subsection{Approximate theory for small mask (``quasi plane-wave'' masked wavefunction)}
\label{app:supp-approximate}

Let us consider first a limiting case for which we are able to obtain simple analytical approximate results that can be easily interpreted. Specifically, we assume that ${w}_{M}\ll {d}_{M}\ll {w}_{0}$, that is the mask radius is very small, and the mask transmission area is located close to the central vortex, entirely on one side of it. With these approximations, we can neglect the variations in the amplitude factor ${r}^{\ell }{e}^{-r^2/w_0^2}$ across the mask transmission area and replace this factor with its value at the mask center. Moreover, in the phase factor ${e}^{\pm i\ell \phi }$ we can use the following approximations:
\begin{equation}
\phi =\arctan\left(\frac{y}{x}\right)\approx \arctan\left(\frac{y}{{x}_{M}}\right)\approx \frac{y}{{d}_{M}}.
\end{equation}
We thus obtain
\begin{equation}
\psi_T=M\left(\bm{r}\right)\psi \left(\bm{r}\right)\approx {N}_{\ell }\frac{{d}_{M}^{\ell }}{{w}_{0}^{\ell +1}}{e}^{-\frac{{d}_{M}^{2}}{{w}_{0}^{2}}}{e}^{-\frac{{\left(x-{d}_{M}\right)}^{2}+{y}^{2}}{{w}_{M}^{2}}}{e}^{\pm i{k}_{s}y}=\frac{{N}_{\ell M}}{{w}_{0}}{e}^{-\frac{{\left(x-{d}_{M}\right)}^{2}+{y}^{2}}{{w}_{M}^{2}}}{e}^{\pm i{k}_{s}y},
\label{eq:supp-1-1}
\end{equation}
where we introduced the \textbf{``superkick'' wavevector} ${k}_{s}=\frac{\ell }{{d}_{M}}$, corresponding to a transverse momentum ${q}_{s}=\hbar {k}_{s}=\frac{\hbar \ell }{{d}_{M}}$, and a new amplitude constant ${N}_{\ell M}$ given by
\begin{equation}
{N}_{\ell M}={\left(\frac{{d}_{M}}{{w}_{0}}\right)}^{\ell }\sqrt{\frac{{2}^{\ell +1}}{\pi \left(\ell !\right)}}{e}^{-\frac{{d}_{M}^{2}}{{w}_{0}^{2}}}.
\label{eq:supp-1-2}
\end{equation}

The masked wavefunction, within this approximation, is simply a Gaussian wave-packet with the amplitude profile corresponding to the mask transmission function and having transverse average wavevector ${k}_{s}$ directed along the $y$ axis (positive or negative depending on the sign of initial OAM). The mean transverse momentum is hence ${q}_{s}$ (also directed along the \textit{y} axis), while the transverse momentum distribution width is now ${w}_{Mq}=2\hbar /{w}_{M}$, which in our approximate limit is much larger than the initial ${w}_{q}$. More precisely, the momentum wavefunction obtained by taking the Fourier transform of \eqref{eq:supp-1-1} is the following:
\begin{equation}
\widetilde{\psi}_T\left(\bm{q}\right)=\frac{{w}_{q}{N}_{\ell M}}{{w}_{Mq}^{2}}{e}^{-\frac{{q}_{x}^{2}+{\left({q}_{y}-{q}_{s}\right)}^{2}}{{w}_{Mq}^{2}}-i\frac{{d}_{M}}{\hbar }{q}_{x}}.
\label{eq:supp-1-3}
\end{equation}

The (unnormalized) momentum distribution of the masked wavefunction in this plane-wave approximation is then given by
\begin{equation}
{\left\lvert \widetilde{\psi}_T\left(\bm{q}\right)\right\rvert}^{2}=\frac{{w}_{q}^{2}{N}_{\ell M}^{2}}{{w}_{Mq}^{4}}{e}^{-2\frac{{q}_{x}^{2}+{\left({q}_{y}-{q}_{s}\right)}^{2}}{{w}_{Mq}^{2}}}.
\label{eq:supp-1-4}
\end{equation}

The increase in momentum distribution width is due to the interaction with the mask and is hence compensated by the random momentum acquired by the mask itself. However, as demonstrated in the Appendix~\ref{app:supp-mask-recoil}, the average momentum ${\bm{q}}_{\bm{s}}$ gained by the photon after the mask is not due to the interaction with the mask, and it can also be much greater than the initial momentum distribution ${w}_{q}$ of the photon (although in our approximation it will usually be smaller than ${w}_{Mq}$, unless $\ell$ is very large). This effect is due to the super-oscillation phenomenon and leads to a local violation of the momentum conservation law in certain single quantum events.

\subsection{Exact mask theory}
\label{app:supp-exact}

From the convolution theorem of Fourier transforms, the effect of the mask product in momentum space can be given as a 2D convolution of $\widetilde{M}\left(\bm{q}\right)$ and $\widetilde{\psi }(\bm{q})$ (divided by a factor $2\pi \hbar$). Hence, after the mask we obtain the following momentum wavefunction:
\begin{equation}
\widetilde{\psi}_T(\bm q)
=(-i)^\ell\frac{N_\ell}{\pi w_q^{\ell+1}w_{Mq}^2}
\iint d^2q'\,
e^{-\frac{(q_x-q_x')^2+(q_y-q_y')^2}{w_{Mq}^2}}
e^{-i\frac{(q_x-q_x')d_M}{\hbar}} (q_x'\pm iq_y')^\ell
e^{-\frac{q_x'^2+q_y'^2}{w_q^2}} .
\label{eq:supp-2-1}
\end{equation}

Now we compute this convolution analytically. For this purpose, it is convenient to replace the ${\left({q}_{x}^{'}\pm i{q}_{y}^{'}\right)}^{\ell }$ factor with derivatives on the subsequent Gaussian, as follows:
\begin{equation}
\begin{aligned}
\widetilde{\psi}_T(\bm q)
&=(-i)^\ell\frac{N_\ell}{\pi w_q^{\ell+1}w_{Mq}^2}
\left(-\frac{w_q^2}{2}\right)^\ell
\iint d^2q'\,
e^{-\frac{(q_x-q_x')^2+(q_y-q_y')^2}{w_{Mq}^2}}
e^{-i\frac{(q_x-q_x')d_M}{\hbar}} 
\left(\partial_{q_x'}\pm i\partial_{q_y'}\right)^\ell
e^{-\frac{q_x'^2+q_y'^2}{w_q^2}} \\
&=i^\ell\frac{N_\ell w_q^{\ell-1}}{\pi 2^\ell w_{Mq}^2}
\iint d^2q'\,
e^{-\frac{(q_x-q_x')^2+(q_y-q_y')^2}{w_{Mq}^2}}
e^{-i\frac{(q_x-q_x')d_M}{\hbar}}
\left(\partial_{q_x'}\pm i\partial_{q_y'}\right)^\ell
e^{-\frac{q_x'^2+q_y'^2}{w_q^2}} .
\end{aligned}
\end{equation}
Then we do an integration by parts on the derivatives, to move them on the other factor within the integral (with a suitable sign change):
\begin{equation}
\widetilde{\psi}_T(\bm q)
=(-i)^\ell\frac{N_\ell w_q^{\ell-1}}{\pi 2^\ell w_{Mq}^2}
\iint d^2q'\,
\left[
\left(\partial_{q_x'}\pm i\partial_{q_y'}\right)^\ell
e^{-\frac{(q_x-q_x')^2+(q_y-q_y')^2}{w_{Mq}^2}}
e^{-i\frac{(q_x-q_x')d_M}{\hbar}}
\right] e^{-\frac{q_x'^2+q_y'^2}{w_q^2}} .
\end{equation}
Because the differentiated factor is symmetric under the interchange of primed and unprimed \textit{q} variables, we can move the derivatives to the unprimed ones (with another sign change) and take them outside the integral:
\begin{equation}
\widetilde{\psi}_T(\bm q)
=i^\ell\frac{N_\ell w_q^{\ell-1}}{\pi 2^\ell w_{Mq}^2}
\left(\partial_{q_x}\pm i\partial_{q_y}\right)^\ell
\iint d^2q'\,
e^{-\frac{(q_x-q_x')^2+(q_y-q_y')^2}{w_{Mq}^2}}
e^{-i\frac{(q_x-q_x')d_M}{\hbar}}
e^{-\frac{q_x'^2+q_y'^2}{w_q^2}} .
\end{equation}
The remaining integral can now be carried out analytically:
\begin{equation}
\begin{aligned}
\widetilde{\psi}_T(\bm q)
&=i^\ell\frac{N_\ell w_q^{\ell-1}}{\pi 2^\ell w_{Mq}^2}
\left(\partial_{q_x}\pm i\partial_{q_y}\right)^\ell
\left[
\pi\frac{w_q^2w_{Mq}^2}{w_q^2+w_{Mq}^2}
e^{-\frac{q_x^2+q_y^2}{w_q^2+w_{Mq}^2}
-i\frac{d_Mw_{Mq}^2q_x}{\hbar(w_q^2+w_{Mq}^2)}
-\frac{d_M^2w_q^2w_{Mq}^2}{4\hbar^2(w_q^2+w_{Mq}^2)}}
\right] \\
&=i^\ell\frac{N_\ell w_q^{\ell+1}}
{2^\ell\left(w_q^2+w_{Mq}^2\right)}
\left(\partial_{q_x}\pm i\partial_{q_y}\right)^\ell
\left[
e^{-\frac{q_x^2+q_y^2}{w_q^2+w_{Mq}^2}
-i\frac{d_Mw_{Mq}^2q_x}{\hbar(w_q^2+w_{Mq}^2)}
-\frac{d_M^2w_q^2w_{Mq}^2}{4\hbar^2(w_q^2+w_{Mq}^2)}}
\right].
\end{aligned}
\end{equation}
Taking the derivatives, we then obtain our main result:
\begin{equation}
\widetilde{\psi}_T(\bm q)
=(-i)^\ell
\frac{N_\ell w_q^{\ell+1}}
{\left(w_q^2+w_{Mq}^2\right)^{\ell+1}}
e^{-\frac{d_M^2w_q^2w_{Mq}^2}
{4\hbar^2\left(w_q^2+w_{Mq}^2\right)}}
\left(q_x+i\frac{d_Mw_{Mq}^2}{2\hbar}\pm iq_y\right)^\ell
e^{-\frac{q_x^2+q_y^2}{w_q^2+w_{Mq}^2}
-i\frac{d_Mw_{Mq}^2q_x}
{\hbar\left(w_q^2+w_{Mq}^2\right)}}.
\label{eq:supp-2-2}
\end{equation}
The probability distribution of the photon momentum after the mask is given by the absolute square of the latter:
\begin{equation}
\begin{aligned}
f_T(\bm q)
&={\left\lvert\widetilde{\psi}_T(\bm q)\right\rvert}^{2} \\
&=\frac{N_\ell^2w_q^{2\ell+2}}
{\left(w_q^2+w_{Mq}^2\right)^{2\ell+2}}
e^{-\frac{d_M^2w_q^2w_{Mq}^2}
{2\hbar^2\left(w_q^2+w_{Mq}^2\right)}}
\left[
q_x^2+\left(\frac{d_Mw_{Mq}^2}{2\hbar}\pm q_y\right)^2
\right]^\ell
e^{-\frac{2(q_x^2+q_y^2)}{w_q^2+w_{Mq}^2}} .
\end{aligned}
\label{eq:supp-2-3}
\end{equation}

This distribution is not normalized. More precisely, its integral returns the transmission probability $P_{\mathrm{tr}}$, as it incorporates the transmission effect of the mask. Hence, the normalized version $f_{T,\mathrm{N}}\left(\bm{q}\right)$ can be obtained easily by dividing \eqref{eq:supp-2-3} by $P_{\mathrm{tr}}$, that is $f_{T,\mathrm{N}}\left(\bm{q}\right)=f_T\left(\bm{q}\right)/P_{\mathrm{tr}}$.

The maximum of $f_T\left(\bm{q}\right)$ for varying ${q}_{y}$ and ${q}_{x}=0$ can be obtained analytically and is as follows:
\begin{equation}
{q}_{y,max}=\pm {w}_{q}\frac{{w}_{0}^{2}}{{w}_{M}^{2}}\left[-\frac{{d}_{M}}{2{w}_{0}}+\sqrt{{\left(\frac{{d}_{M}}{2{w}_{0}}\right)}^{2}+\frac{\ell {w}_{M}^{2}}{2{w}_{0}^{2}}\left(1+\frac{{w}_{M}^{2}}{{w}_{0}^{2}}\right)}\right].
\label{eq:supp-2-4}
\end{equation}
If ${w}_{M}\ll {d}_{M}\ll {w}_{0}$ we obtain the following approximation:
\begin{equation}
{q}_{y,max}\approx \pm {w}_{q}\frac{{w}_{0}\ell }{2{d}_{M}}=\pm \frac{\hbar \ell }{{d}_{M}},
\label{eq:supp-2-5}
\end{equation}
which corresponds to the superkick predicted in the approximate theory. Thus, we confirm with this exact theory that a shifted momentum distribution appears that can exhibit the superkick phenomenon. However, the exact theory shows a superkick whenever ${w}_{M}\ll {w}_{0}$ and ${d}_{M}\ll {w}_{0}$, with no need for imposing also ${w}_{M}\ll {d}_{M}$. Moreover, it is evident from \eqref{eq:supp-2-4} that the superkick grows without limits for increasing $\ell$.

As we found numerically, in general the superkick effect, in particular for small $\ell$, is optimized when ${w}_{M}\sim {d}_{M}$ (although ${q}_{y,max}$ is not maximized). More precisely, we find that depending on the exact mask parameters, setting ${w}_{M}$ in the range ${d}_{M}$--$2{d}_{M}$ is optimal for experimental verification of the effect, considering also the mask transmission probability (which decreases with decreasing ${w}_{M}$ and ${d}_{M}$).

It is also interesting to compute the mean value of ${q}_{y}$:
\begin{equation}
\bar q_{Ty}={\left\langle {q}_{y}\right\rangle}_{f_{T,\mathrm{N}}}=\iint_{-\infty }^{+\infty } {q}_{y}f_{T,\mathrm{N}}\,d{q}_{x}d{q}_{y}.
\end{equation}
We did not look for a general analytic form (although it might be possible that one exists). We only report the case $\ell =1$, for which the computation is simple:
\begin{equation}
\bar q_{Ty}
={\left\langle q_y\right\rangle}_{f_{T,\mathrm N}}
=\frac{\pm\hbar/d_M}
{\dfrac{w_{Mq}^2}{w_q^2+w_{Mq}^2}
+2\dfrac{\hbar^2}{w_{Mq}^2d_M^2}}
=\frac{\pm q_s}
{\left(1+\dfrac{w_q^2}{w_{Mq}^2}\right)^{-1}
+2\dfrac{q_s^2}{w_{Mq}^2}}
=\frac{\pm\hbar/d_M}
{\dfrac{w_0^2}{w_0^2+w_M^2}
+\dfrac{w_M^2}{2d_M^2}} .
\end{equation}
In the limit ${w}_{M}\ll {d}_{M}\ll {w}_{0}$ we recover the expected $\bar q_{Ty}=\pm {q}_{s}=\pm \frac{\hbar }{{d}_{M}}$. If instead we have ${w}_{M}\simeq {d}_{M}\ll {w}_{0}$, we have ${w}_{Mq}\simeq 2{q}_{s}$ and we obtain $\bar q_{Ty}\approx \pm \frac{2}{3}{q}_{s}$, that is the superkick effect is slightly reduced (but more easily observable, as ${w}_{M}$ needs not be as small).

It should be noted that the average variation of momentum for the photons that pass the mask is balanced by the average variation of momentum for the photons which do not pass the mask. Indeed, by an explicit calculation for $\ell =1$ we find:
\begin{equation}
P_{\mathrm{tr}}\bar q_{Ty}=\iint_{-\infty }^{+\infty } {q}_{y}f_Td{q}_{x}d{q}_{y}=\pm \frac{2{d}_{M}{w}_{M}^{2}\hbar }{{\left({w}_{0}^{2}+{w}_{M}^{2}\right)}^{2}}{e}^{-\frac{2{d}_{M}^{2}}{{w}_{0}^{2}+{w}_{M}^{2}}}.
\end{equation}

At the same time, we may compute explicitly the average \textit{y} momentum of the photon in the case it is not transmitted (N) by the mask (e.g. it could be reflected) in the limit of small mask transmission amplitude and we find
\begin{equation}
\left(1-P_{\mathrm{tr}}\right){q}_{Ny,mean}=\iint_{-\infty }^{+\infty } {q}_{y}{f}_{N}d{q}_{x}d{q}_{y}\approx \mp \frac{2{d}_{M}{w}_{M}^{2}\hbar }{{\left({w}_{0}^{2}+{w}_{M}^{2}\right)}^{2}}{e}^{-\frac{2{d}_{M}^{2}}{{w}_{0}^{2}+{w}_{M}^{2}}}
\end{equation}
(the derivation of the non-transmitted-branch wavefunction $\Psi_N$, from which the associated momentum distribution $f_N$ can be derived, is reported in Appendix \ref{app:supp-mask-recoil}; in $\Psi_N$ the mask motion is also included, but by setting $w_C=0$ the mask becomes fixed and the photon-only wavefunction $\psi_N$ is obtained).

Hence, the mask does not induce any variation in the total ensemble average momentum of the photon. This average momentum balance ensures that the superkick in a macroscopic situation cannot induce a local violation of conservation laws. However, for individual quantum events for which the photon has been transmitted the momentum is locally not conserved, as the superkick momentum computed above is not exchanged with the mask for transmitted photons.

\subsection{Reference null-hypothesis distributions: unbiased local momentum-conserving (ULMC) and maximal local momentum-conserving (MLMC) theories}
\label{app:supp-null-hypotheses}

In this subsection we construct the reference null-hypothesis distributions against the measured distribution is compared to demonstrate the existence of the superkick phenomenon and the corresponding momentum local non-conservation.

A first possible reference is as follows. Assume that the mask-induced momentum kick, drawn randomly from the mask momentum distribution, is just added to the whole input momentum distribution of the input LG mode. The resulting normalized momentum probability distribution would then be the convolution of the initial photon momentum distribution \eqref{eq:supp-0-5} and the normalized mask-kick distribution \eqref{eq:supp-0-9}:
\begin{equation}
f_R(\bm q)
={\left\lvert\widetilde{\psi}(\bm q)\right\rvert}^{2}
\ast K_M(\bm q)
=\frac{2N_\ell^2}{\pi w_q^{2\ell+2}w_{Mq}^2}
\iint d^2q'\,
e^{-2\frac{(q_x-q_x')^2+(q_y-q_y')^2}{w_{Mq}^2}}
\left(q_x'^2+q_y'^2\right)^\ell
e^{-2\frac{q_x'^2+q_y'^2}{w_q^2}} .
\label{eq:supp-3-1}
\end{equation}
Using similar tricks as for the previous calculation, we obtain:
\begin{equation}
\begin{aligned}
f_R(\bm q)
&=\left.
\frac{2N_\ell^2}{\pi w_q^{2\ell+2}w_{Mq}^2}
\iint d^2q'\,
e^{-2\frac{(q_x-q_x')^2+(q_y-q_y')^2}{w_{Mq}^2}}
\left(q_x'^2+q_y'^2\right)^\ell
e^{-2\xi\frac{q_x'^2+q_y'^2}{w_q^2}}
\right|_{\xi=1} \\
&=\left.
N_\ell^2\left(-\frac12\right)^\ell
\frac{1}{w_{Mq}^2}
\frac{\partial^\ell}{\partial\zeta^\ell}
\left[
\frac{1}{\zeta}
e^{-2\frac{q_x^2+q_y^2}{w_{Mq}^2}
\left(1-\frac{w_q^2}{\zeta w_{Mq}^2}\right)}
\right]
\right|_{\zeta=1+\frac{w_q^2}{w_{Mq}^2}} .
\end{aligned}
\label{eq:supp-3-2}
\end{equation}
Here $\zeta=\xi+w_q^2/w_{Mq}^2$, so that differentiation with respect to $\zeta$ is equivalent to differentiation with respect to the auxiliary parameter $\xi$.

It is evident that this result is azimuthally symmetric around the origin of momenta, hence it corresponds to a vanishing average ${q}_{y}$ momentum. The width of the distribution is controlled by the Gaussian term which has a width $\sqrt{{w}_{q}^{2}+{w}_{Mq}^{2}}$, but the derivatives also generate a polynomial factor that makes the exact expression of the width more complex. We name this distribution ``\textit{unbiased local momentum conserving}'' (ULMC) theory.

This is however not the most meaningful reference distribution we have to compare with. Indeed, even in a model with no superkick, it is quite reasonable to assume that the mask, before adding its random momentum impulse, will somehow select (filter) only a subset of the momenta of the input beam. The fraction of the input momenta distribution that the mask can select is constrained by the fact that the overall mask transmission probability is fixed. In other words, the mask-selected distribution of momenta of the input beam must obey the following equation:
\begin{equation}
\iint {S}_{M}\left(\bm{q}\right){\left\lvert \widetilde{\psi }\left(\bm{q}\right)\right\rvert}^{2}d{q}^{2}=P_{\mathrm{tr}},
\label{eq:supp-3-3}
\end{equation}
where ${S}_{M}\left(\bm{q}\right)$ is the hypothesized mask-selection probability for each momentum. If we assume that the mask selects only the highest possible \textit{y}-oriented momenta (for positive OAM and $x_M$), to get as close as possible to the superkick case, one has
\begin{equation}
{S}_{M}\left({q}_{x},{q}_{y}\right)=\theta \left({q}_{y}-{q}_{0}\right)
\label{eq:supp-3-4},
\end{equation}
where $\theta$ is the Heaviside step function and ${q}_{0}$ is a threshold \textit{y}-oriented momentum above which the mask transmits the photons, to be determined from the condition
\begin{equation}
\int_{{q}_{0}}^{+\infty } d{q}_{y}\int_{-\infty }^{+\infty } d{q}_{x}{\left\lvert \widetilde{\psi }\left(\bm{q}\right)\right\rvert}^{2}=P_{\mathrm{tr}}.
\label{eq:supp-3-5}
\end{equation}

Once ${q}_{0}$ is determined, the normalized reference distribution is given again by a convolution with the mask momentum distribution (indeed, there is no reason to believe that the mask random momentum impulse is biased owing to the modified input momentum distribution, since the mask affects equally all input plane waves). Following the main-text notation, define the normalized selected upper-tail distribution as
\begin{equation}
f_S(\bm q)=N_S f(\bm q)\theta(q_y-q_0),\qquad N_S=\frac{1}{P_{\mathrm{tr}}}.
\end{equation}

\begin{equation}
\begin{aligned}
f_{SM}(\bm q)
&=f_S(\bm q)\ast K_M(\bm q)
=\left[
\frac{{\left\lvert\widetilde{\psi}(\bm q)\right\rvert}^{2}
\theta(q_y-q_0)}
{P_{\mathrm{tr}}}
\right]\ast K_M(\bm q) \\
&=\frac{2N_\ell^2}
{\pi P_{\mathrm{tr}}w_q^{2\ell+2}w_{Mq}^2}
\iint d^2q'\,
e^{-2\frac{(q_x-q_x')^2+(q_y-q_y')^2}{w_{Mq}^2}}
\theta(q_y'-q_0)
\left(q_x'^2+q_y'^2\right)^\ell
e^{-2\frac{q_x'^2+q_y'^2}{w_q^2}} .
\end{aligned}
\label{eq:supp-3-6}
\end{equation}
This distribution will be named \textit{maximal local momentum-conserving} distribution (MLMC) and will represent our main null hypothesis used to statistically test the violation of local momentum conservation in the measured distribution.

Although expression \eqref{eq:supp-3-6} can be computed analytically by Wolfram Mathematica at least for $\ell =1$ (in terms of error functions), it is rather complex and computationally heavy. It is always possible to compute it numerically, as we do for our data analysis procedure. Alternatively, since we are mainly interested in cases in which $P_{\mathrm{tr}}$ is very small and hence ${q}_{0}$ will bound the very small tail of the input-beam momentum distribution in the \textit{y} direction, we can usually adopt the approximation
\begin{equation}
{\left\lvert \widetilde{\psi }\left(\bm{q}\right)\right\rvert}^{2}\theta \left({q}_{y}-{q}_{0}\right)\approx P_{\mathrm{tr}}\delta \left({q}_{x}\right)\delta \left({q}_{y}-\bar q_{Sy}\right),
\label{eq:supp-3-7}
\end{equation}
where $\bar q_{Sy}$ is the mean ${q}_{y}$-momentum corresponding to the distribution ${\left\lvert \widetilde{\psi }\left(\bm{q}\right)\right\rvert}^{2}\theta \left({q}_{y}-{q}_{0}\right)$, that can be computed as follows:
\begin{equation}
\bar q_{Sy}=\frac{1}{P_{\mathrm{tr}}}\int_{{q}_{0}}^{+\infty } d{q}_{y}{q}_{y}\int_{-\infty }^{+\infty } d{q}_{x}{\left\lvert \widetilde{\psi }\left(\bm{q}\right)\right\rvert}^{2}.
\label{eq:supp-3-8}
\end{equation}
For $\ell =1$, this can also be computed analytically as a function of ${q}_{0}$ and is given by (as computed by Mathematica):
\begin{equation}
\bar q_{Sy}=\frac{1}{2\sqrt{2\pi }}\frac{{w}_{0}^{5}{\left(1+\frac{{w}_{M}^{2}}{{w}_{0}^{2}}\right)}^{3}\left({q}_{0}^{2}+3\frac{{\hbar }^{2}}{{w}_{0}^{2}}\right)}{{w}_{M}^{2}\left[2{d}_{M}^{2}+{w}_{M}^{2}\left(1+\frac{{w}_{M}^{2}}{{w}_{0}^{2}}\right)\right]\hbar }{e}^{\frac{2{d}_{M}^{2}}{{w}_{0}^{2}+{w}_{M}^{2}}-\frac{{q}_{0}^{2}{w}_{0}^{2}}{2{\hbar }^{2}}}.
\end{equation}
Using approximation \eqref{eq:supp-3-7}, we then get
\begin{equation}
f_{SM}\left(\bm{q}\right)\approx \left[\delta \left({q}_{x}\right)\delta ({q}_{y}-\bar q_{Sy})\right]\ast K_M(\bm q)=\frac{2}{\pi {w}_{Mq}^{2}}{e}^{-2\frac{{q}_{x}^{2}+{\left({q}_{y}-\bar q_{Sy}\right)}^{2}}{{w}_{Mq}^{2}}}.
\label{eq:supp-3-9}
\end{equation}

\section{Modeling the mask-photon interaction and mask recoil}
\label{app:supp-mask-recoil}

We model the following argument after that given by Yakir Aharonov, Sandu Popescu and Daniel Rohrlich as published in \cite{aharonov_conservation_2023} and further refined by Daniel Collins and Sandu Popescu \cite{collins_conservation_2025}. It is adapted to move from angular momentum to (linear) momentum conservation, since this is what we are doing experimentally, and to make it much closer to describing our actual experimental situation.

We consider a photon travelling in free space towards the spatial region where the mask is located and given by the wavefunction ${\psi }_{Ph}$ described by Eq.~\eqref{eq:supp-0-1} (we are inserting a ``\textit{Ph}'' label for greater clarity since we are going to introduce a second quantum object, that is the mask). As in the previous sections, we will omit the $z$ and $t$ variables for brevity. The corresponding transverse momentum space OAM wavefunction is given in \eqref{eq:supp-0-3}. The momentum probability distribution is given by Eq.~\eqref{eq:supp-0-5}.

The transverse momentum of the photon in free space is a conserved quantity, hence Eq.~\eqref{eq:supp-0-5} will be valid for all values of time $t$ and longitudinal position $z$, until an interaction occurs. Vortex wavefunctions as \eqref{eq:supp-0-1} have a super-oscillatory behavior in the central region, close to the vortex. Hence, we expect that, if we ``extract'' a photon from that region, it may exhibit anomalous values of transverse momentum that are not predicted by Eq.~\eqref{eq:supp-0-5} and which we call ``superkick'' momenta.

Let us now also model the mask used to probe the super-oscillatory region of the input photon as a quantum object $\mathcal M$ (possibly with very large mass, but it is not strictly necessary). We describe the position ${\bm{r}}_{C}$ of the center of mass (C) of $\mathcal M$ with a wavefunction initially given by
\begin{equation}
\psi_C\left({\bm{r}}_{C}\right)=\frac{1}{{w}_{C}}\sqrt{\frac{2}{\pi }}{e}^{-\frac{{\left({x}_{C}-{d}_{M}\right)}^{2}+{y}_{C}^{2}}{{w}_{C}^{2}}},
\label{eq:supp-4-1}
\end{equation}
corresponding to a Gaussian function centered at distance ${d}_{M}$ from the origin and of width ${w}_{C}$. For definiteness, we have placed the mask center on the positive \textit{x} semiaxis, so that it has mean coordinates $\langle{x}_{C}\rangle={d}_{M}, \langle{y}_{C}\rangle=0$. We also attribute to the system (photon + mask) an additional internal degree of freedom (a ``pointer''), which we model as a two-state quantum degree of freedom with eigenstates $\left\lvert {\left.T\right\rangle}_{P}\right.$ and $\left\lvert {\left.N\right\rangle}_{P}\right.$, which stand for ``\textit{transmitted}'' and ``\textit{not transmitted}'', respectively. If the system remains in state $\left\lvert {\left.N\right\rangle}_{P}\right.$ the photon is blocked by the mask (e.g. reflected by the mask), while when the system is in state ${\left\lvert \left.T\right\rangle\right.}_{P}$ the photon has been transmitted through the mask. The pointer state can be postselected by detecting if the photon has actually passed the mask, as done in the real experiment (in this case the pointer state can be actually identified as a degree of freedom of the photon itself).

We will also need the mask momentum distribution, which is given by
\begin{equation}
{\left\lvert \widetilde{\psi}_C\left({\bm{q}}_{C}\right)\right\rvert}^{2}=\frac{2}{\pi {w}_{Cq}^{2}}{e}^{-2\frac{{q}_{Cx}^{2}+{q}_{Cy}^{2}}{{w}_{Cq}^{2}}},
\label{eq:supp-4-2}
\end{equation}
where we introduced the mask momentum distribution width ${w}_{Cq}=2\hbar /{w}_{C}$.

The initial quantum state (at, say, $t={0}^{-}$) of the photon+mask system is then taken to be the following:
\begin{equation}
\left\lvert \left.{\Psi }_{0}\right\rangle\right.={\psi }_{Ph}\left(\bm{r}\right){\psi }_{C}\left({\bm{r}}_{C}\right)\left\lvert {\left.N\right\rangle}_{P}\right. .
\end{equation}

The photon-mask interaction Hamiltonian we assume is the following:
\begin{equation}
{\widehat{H}}_{I}=-\hbar \delta \left(t\right)g\left(x-{x}_{C},y-{y}_{C}\right){\widehat{\sigma }}_{1P},
\end{equation}
where ${\widehat{\sigma }}_{1P}$ is the first Pauli matrix acting on the P degree of freedom, represented as follows
\begin{equation}
{\widehat{\sigma }}_{1P}=\left(\begin{matrix}0 & 1 \\ 1 & 0\end{matrix}\right),
\end{equation}
and $g$ is a dimensionless \textit{interaction distribution} (correlated with the spatial extension of the mask) which we take to be given by the following expression:
\begin{equation}
g\left(\Delta x,\Delta y\right)=\arcsin\left[{A}_{g}{e}^{-\frac{{\Delta x}^{2}+\Delta {y}^{2}}{{w}_{M}^{2}}}\right],
\end{equation}
where ${w}_{M}$ is the mask width. In actual experimental implementations where the mask is macroscopic, one has ${w}_{C}\ll {w}_{M}$, but this is not needed for the validity of the following. We also assume that the amplitude constant ${A}_{g}<1$, so that the arcsin function is real and single-valued. The temporal behavior of the interaction is here modeled for simplicity with the Dirac delta function $\delta \left(t\right)$, but a finite-duration Hamiltonian would give similar results.

The evolution operator over a short interval $\tau$ containing $t=0$ is given by (the free particle Hamiltonians can be neglected due to the impulsive nature of the interaction):
\begin{equation}
\begin{aligned}
\widehat U(0;\tau)
&=e^{-\frac{i}{\hbar}\int_0^\tau \widehat H_I(t)\,dt} 
=e^{ig(x-x_C,y-y_C)\widehat\sigma_{1P}} \\
&=\widehat I_P\cos\!\left[g(x-x_C,y-y_C)\right]
+i\widehat\sigma_{1P}\sin\!\left[g(x-x_C,y-y_C)\right] \\
&=\widehat I_P
\sqrt{1-A_g^2e^{-2\frac{(x-x_C)^2+(y-y_C)^2}{w_M^2}}}
+i\widehat\sigma_{1P}A_g
e^{-\frac{(x-x_C)^2+(y-y_C)^2}{w_M^2}} ,
\end{aligned}
\end{equation}
where ${\widehat{I}}_{P}$ is the identity operator in the pointer space. Hence, the system quantum state after the mask is as follows
\begin{equation}
\begin{aligned}
\lvert\Psi_1\rangle
&=\widehat U(0;\tau)\lvert\Psi_0\rangle \\
&=\sqrt{1-A_g^2e^{-2\frac{(x-x_C)^2+(y-y_C)^2}{w_M^2}}}\,
\psi_{Ph}(\bm r)\psi_C(\bm r_C)\lvert N\rangle_P \\
&\quad+iA_g e^{-\frac{(x-x_C)^2+(y-y_C)^2}{w_M^2}}\,
\psi_{Ph}(\bm r)\psi_C(\bm r_C)\lvert T\rangle_P .
\end{aligned}
\end{equation}

If we measure the pointer degree of freedom in the \textit{N/T} basis, we will obtain the following two wavefunctions for blocked and transmitted photons (neglecting a global phase):
\begin{equation}
\begin{aligned}
\Psi_N
&=\sqrt{1-A_g^2e^{-2\frac{(x-x_C)^2+(y-y_C)^2}{w_M^2}}}\,
\psi_{Ph}(\bm r)\psi_C(\bm r_C) \\
&=\frac{N_\ell}{w_0^{\ell+1}w_C}\sqrt{\frac{2}{\pi}}\,
\sqrt{1-A_g^2e^{-2\frac{(x-x_C)^2+(y-y_C)^2}{w_M^2}}}\,
(x\pm iy)^\ell
e^{-\frac{x^2+y^2}{w_0^2}}
e^{-\frac{(x_C-d_M)^2+y_C^2}{w_C^2}} .
\end{aligned}
\end{equation}

\begin{equation}
\begin{aligned}
\Psi_T
&=A_g e^{-\frac{(x-x_C)^2+(y-y_C)^2}{w_M^2}}\,
\psi_{Ph}(\bm r)\psi_C(\bm r_C) \\
&=\frac{2A_g}{\pi w_0^{\ell+1}w_C}
\sqrt{\frac{2^\ell}{\ell!}}\,(x\pm iy)^\ell
e^{-\frac{(x-x_C)^2+(y-y_C)^2}{w_M^2}
-\frac{x^2+y^2}{w_0^2}
-\frac{(x_C-d_M)^2+y_C^2}{w_C^2}} \\
&=\frac{2A_g e^{-d_M^2/w_C^2}}{\pi w_0^{\ell+1}w_C}
\sqrt{\frac{2^\ell}{\ell!}}\,(x\pm iy)^\ell
e^{-\frac{x^2+y^2}{w_{0M}^2}
-\frac{x_C^2+y_C^2}{w_{CM}^2}
+2\frac{xx_C+yy_C}{w_M^2}
+2\frac{d_Mx_C}{w_C^2}} ,
\end{aligned}
\end{equation}
where we introduced
\begin{equation}
{w}_{0M}=\frac{{w}_{0}{w}_{M}}{\sqrt{{w}_{0}^{2}+{w}_{M}^{2}}}, {w}_{CM}=\frac{{w}_{C}{w}_{M}}{\sqrt{{w}_{C}^{2}+{w}_{M}^{2}}} ,
\end{equation}
or equivalently we set:
\begin{equation}
\frac{1}{{w}_{0M}^{2}}=\frac{1}{{w}_{0}^{2}}+\frac{1}{{w}_{M}^{2}}, \frac{1}{{w}_{CM}^{2}}=\frac{1}{{w}_{C}^{2}}+\frac{1}{{w}_{M}^{2}}.
\end{equation}

The transmitted photon wavefunction ${\Psi }_{T}$ is not normalized and its integral gives the transmission probability. The general expression for this is as follows:
\begin{equation}
\begin{aligned}
P_{\mathrm{tr}}
&=\frac{A_g^2w_M^2\left(w_C^2+w_M^2\right)^\ell}
{\left(w_0^2+w_C^2+w_M^2\right)^{\ell+1}}
e^{-\frac{2d_M^2}{w_C^2+w_M^2}}
{}_1\mathrm F_1\!\left(
\ell+1;1;
\frac{2d_M^2w_0^2}
{\left(w_C^2+w_M^2\right)\left(w_0^2+w_C^2+w_M^2\right)}
\right) \\
&=\frac{A_g^2w_M^2\left(w_C^2+w_M^2\right)^\ell}
{\left(w_0^2+w_C^2+w_M^2\right)^{\ell+1}}
e^{-\frac{2d_M^2}{w_C^2+w_M^2}}
L_{-\ell-1}\!\left[
\frac{2d_M^2w_0^2}
{\left(w_C^2+w_M^2\right)\left(w_0^2+w_C^2+w_M^2\right)}
\right],
\end{aligned}
\end{equation}
where ${{}_{1}^{}\mathrm{F}}_{1}$ is the Kummer confluent hypergeometric function and ${L}_{n}$ are the Laguerre polynomials. In the specific case $\ell =1$, this simplifies into
\begin{equation}
P_{\mathrm{tr}}(\ell =1)=\frac{{A}_{g}^{2}{w}_{M}^{2}\left[2{d}_{M}^{2}{w}_{0}^{2}+\left({w}_{C}^{2}+{w}_{M}^{2}\right)\left({w}_{0}^{2}+{w}_{C}^{2}+{w}_{M}^{2}\right)\right]}{{\left({w}_{0}^{2}+{w}_{C}^{2}+{w}_{M}^{2}\right)}^{3}}{e}^{-\frac{2{d}_{M}^{2}}{{w}_{0}^{2}+{w}_{C}^{2}+{w}_{M}^{2}}}.
\end{equation}

${\Psi }_{T}$ can be normalized after dividing by $\sqrt{P_{\mathrm{tr}}}$. These results reduce to those already reported in the previous Sections in the limit ${w}_{C}\to 0$ valid for a classical mask that is not moving.

Now we can Fourier-transform ${\Psi }_{T}$ to obtain the wavefunction and probability distribution for all the transverse momenta:
\begin{equation}
\begin{aligned}
\widetilde\Psi_T
&=\frac{1}{(2\pi\hbar)^2}
\frac{2A_g e^{-d_M^2/w_C^2}}{\pi w_0^{\ell+1}w_C}
\sqrt{\frac{2^\ell}{\ell!}} \\
&\quad\times
\iint dx_C\,dy_C\,
e^{-\frac{x_C^2+y_C^2}{w_{CM}^2}
+2\frac{d_Mx_C}{w_C^2}
-i\frac{q_{Cx}x_C+q_{Cy}y_C}{\hbar}}
\iint dx\,dy\,(x\pm iy)^\ell
e^{-\frac{x^2+y^2}{w_{0M}^2}
+2\frac{xx_C+yy_C}{w_M^2}
-i\frac{q_xx+q_yy}{\hbar}} \\
&=\frac{1}{(2\pi\hbar)^2}
\frac{2A_g e^{-d_M^2/w_C^2}}{\pi w_0^{\ell+1}w_C}
\sqrt{\frac{2^\ell}{\ell!}}\,
i^\ell\hbar^\ell
\left(\partial_{q_x}\pm i\partial_{q_y}\right)^\ell \\
&\quad\times
\iint dx_C\,dy_C\,
e^{-\frac{x_C^2+y_C^2}{w_{CM}^2}
+2\frac{d_Mx_C}{w_C^2}
-i\frac{q_{Cx}x_C+q_{Cy}y_C}{\hbar}}
\iint dx\,dy\,
e^{-\frac{x^2+y^2}{w_{0M}^2}
+2\frac{xx_C+yy_C}{w_M^2}
-i\frac{q_xx+q_yy}{\hbar}} \\
&=\sqrt{\frac{2^\ell}{\ell!}}
\left(\frac{i\hbar}{w_0}\right)^\ell
\left(\partial_{q_x}\pm i\partial_{q_y}\right)^\ell
\widetilde\Psi_{T,\ell=0}.
\end{aligned}
\end{equation}

Let us compute first ${\widetilde{\Psi }}_{T,\ell =0}$:
\begin{equation}
\begin{aligned}
\widetilde\Psi_{T,\ell=0}
&=\frac{1}{(2\pi\hbar)^2}
\frac{2A_g e^{-d_M^2/w_C^2}}{\pi w_0w_C} \\
&\quad\times
\iint dx_C\,dy_C\,
e^{-\frac{x_C^2+y_C^2}{w_{CM}^2}
+2\frac{d_Mx_C}{w_C^2}
-i\frac{q_{Cx}x_C+q_{Cy}y_C}{\hbar}}
\iint dx\,dy\,
e^{-\frac{x^2+y^2}{w_{0M}^2}
+2\frac{xx_C+yy_C}{w_M^2}
-i\frac{q_xx+q_yy}{\hbar}} \\
&=\frac{A_gw_0w_Cw_M^2}
{2\pi\hbar^2\left(w_0^2+w_C^2+w_M^2\right)}
e^{-\frac{d_M^2}{w_0^2+w_C^2+w_M^2}} \\
&\quad\times
e^{
-\frac{
w_C^2(w_0^2+w_M^2)(q_{Cx}^2+q_{Cy}^2)
+w_0^2(w_C^2+w_M^2)(q_x^2+q_y^2)
+2w_0^2w_C^2(q_{Cx}q_x+q_{Cy}q_y)
+4i\hbar d_M\!\left[q_xw_0^2+q_{Cx}(w_0^2+w_M^2)\right]}
{4\left(w_0^2+w_C^2+w_M^2\right)\hbar^2}
}.
\end{aligned}
\end{equation}

Hence, we have
\begin{equation}
\begin{aligned}
\widetilde\Psi_T
&=i^\ell
\frac{A_g\hbar^{\ell-2}w_Cw_M^2}
{2\pi w_0^{\ell-1}\left(w_0^2+w_C^2+w_M^2\right)}
\sqrt{\frac{2^\ell}{\ell!}}\,
e^{-\frac{d_M^2}{w_0^2+w_C^2+w_M^2}}
\left(\partial_{q_x}\pm i\partial_{q_y}\right)^\ell \\
&\quad\times
e^{
-\frac{
w_0^2(w_C^2+w_M^2)(q_x^2+q_y^2)
+w_C^2(w_0^2+w_M^2)(q_{Cx}^2+q_{Cy}^2)
{}+2w_0^2w_C^2(q_{Cx}q_x+q_{Cy}q_y)
+4i\hbar d_M\!\left[q_xw_0^2+q_{Cx}(w_0^2+w_M^2)\right]
}
{4\left(w_0^2+w_C^2+w_M^2\right)\hbar^2}
}.
\end{aligned}
\end{equation}

The general expression is not needed. For $\ell =1$, we obtain
\begin{equation}
\begin{aligned}
\widetilde\Psi_T
&=-i\frac{A_gw_0^2w_Cw_M^2}
{\sqrt8\pi\hbar^3\left(w_0^2+w_C^2+w_M^2\right)^2}
e^{-\frac{d_M^2}{w_0^2+w_C^2+w_M^2}}
\left[
(w_C^2+w_M^2)(q_x\pm iq_y)
+w_C^2(q_{Cx}\pm iq_{Cy})
+2i\hbar d_M
\right] \\
&\quad\times
e^{
-\frac{
w_0^2(w_C^2+w_M^2)(q_x^2+q_y^2)
+w_C^2(w_0^2+w_M^2)(q_{Cx}^2+q_{Cy}^2)
{}+2w_0^2w_C^2(q_{Cx}q_x+q_{Cy}q_y)
+4i\hbar d_M\!\left[q_xw_0^2+q_{Cx}(w_0^2+w_M^2)\right]
}
{4\left(w_0^2+w_C^2+w_M^2\right)\hbar^2}
}.
\end{aligned}
\end{equation}

The corresponding joint momentum distribution is
\begin{equation}
\begin{aligned}
f_T(\bm q,\bm q_C)
&={\left\lvert\widetilde\Psi_T\right\rvert}^{2} \\
&=\frac{A_g^2w_0^4w_C^2w_M^4}
{8\pi^2\hbar^6\left(w_0^2+w_C^2+w_M^2\right)^4}
e^{-\frac{2d_M^2}{w_0^2+w_C^2+w_M^2}} \\
&\quad\times
\left\{
\left[(w_C^2+w_M^2)q_x+w_C^2q_{Cx}\right]^2
+\left[(w_C^2+w_M^2)q_y+w_C^2q_{Cy}\pm2\hbar d_M\right]^2
\right\} \\
&\quad\times
e^{
-\frac{
w_0^2(w_C^2+w_M^2)(q_x^2+q_y^2)
+w_C^2(w_0^2+w_M^2)(q_{Cx}^2+q_{Cy}^2)
{}+2w_0^2w_C^2(q_{Cx}q_x+q_{Cy}q_y)
}
{2\left(w_0^2+w_C^2+w_M^2\right)\hbar^2}
}.
\end{aligned}
\end{equation}

In the limit ${w}_{C}\to 0$, this distribution reduces to that already found for the photon only. Let us now compute the average momenta and other properties predicted by this distribution.

The average momenta for both the photon and the mask along $x$ vanish identically, as can be proved from the distribution symmetry for the transformation ${q}_{x}\to -{q}_{x}, {q}_{Cx}\to -{q}_{Cx}$.

The photon momentum along $y$ instead does not vanish and is given by
\begin{equation}
\begin{aligned}
\left\langle q_y\right\rangle
&=\frac{1}{P_{\mathrm{tr}}}
\iint\!\iint q_y f_T\left(\bm q,\bm q_C\right)
\,d^2q\,d^2q_C \\
&=\pm
\frac{2d_M\left(w_0^2+w_C^2+w_M^2\right)\hbar}
{2d_M^2w_0^2
+\left(w_C^2+w_M^2\right)
\left(w_0^2+w_C^2+w_M^2\right)} \\
&=\pm
\frac{\hbar/d_M}
{\dfrac{w_0^2}{w_0^2+w_C^2+w_M^2}
+\dfrac{w_C^2+w_M^2}{2d_M^2}} .
\end{aligned}
\end{equation}

This is the superkick momentum including the effect of mask recoil (controlled by the distribution width ${w}_{C}$). It diverges in the limit ${d}_{M}\to 0$ if one also ensures that $\sqrt{{w}_{C}^{2}+{w}_{M}^{2}}\sim {d}_{M}\to 0$, showing that the superkick can be ideally very large, irrespective of the original distribution of photon momenta.

The average mask momentum is instead
\begin{equation}
\left\langle {q}_{Cy}\right\rangle=\frac{1}{P_{\mathrm{tr}}}\iint \iint {q}_{Cy}{f}_{T}\left(\bm{q},{\bm{q}}_{C}\right){d}^{2}q{d}^{2}{q}_{C}=0.
\end{equation}

This result alone already shows that the average photon superkick momentum does not originate from the mask, which is our main result here.

The mask however exchanges a \textit{random momentum} with the photon that broadens its momentum distribution. The random momentum distribution is controlled by the mask interaction width ${w}_{M}$ and has typical size $2\hbar /{w}_{M}$.

The photon momentum marginal distribution after the mask is the following:
\begin{equation}
\begin{aligned}
f_T(\bm q)
&=\iint f_T\left(\bm q,\bm q_C\right)\,d^2q_C \\
&=\frac{A_g^2w_0^4w_M^4}
{\pi\left(w_0^2+w_M^2\right)^2
\left(w_0^2+w_C^2+w_M^2\right)\hbar^2}
e^{
-\frac{2d_M^2}{w_0^2+w_C^2+w_M^2}
-\frac{w_0^2w_M^2(q_x^2+q_y^2)}
{2\left(w_0^2+w_M^2\right)\hbar^2}
} \\
&\quad\times
\left[
\frac{w_M^4(q_x^2+q_y^2)}
{4\hbar^2\left(w_0^2+w_M^2\right)}
\pm\frac{d_Mw_M^2q_y}
{\hbar\left(w_0^2+w_C^2+w_M^2\right)}
+\frac{
w_C^2\left(w_0^2+w_C^2+w_M^2\right)
+2d_M^2\left(w_0^2+w_M^2\right)}
{2\left(w_0^2+w_C^2+w_M^2\right)^2}
\right],
\end{aligned}
\end{equation}
which has a broadened width given by $2\hbar /{w}_{0M}>2\hbar /{w}_{0}$. The corresponding marginal distribution for the mask momentum is:
\begin{equation}
f_T(\bm q_C)
=\frac{A_g^2w_C^2w_M^4}
{2\pi\left(w_0^2+w_C^2+w_M^2\right)^2\hbar^2}
e^{
-\frac{2d_M^2}{w_0^2+w_C^2+w_M^2}
-\frac{w_C^2w_M^2(q_{Cx}^2+q_{Cy}^2)}
{2\left(w_C^2+w_M^2\right)\hbar^2}
}
\left[
1+\frac{2d_M^2w_0^2}
{\left(w_C^2+w_M^2\right)
\left(w_0^2+w_C^2+w_M^2\right)}
\right].
\end{equation}

We can see that also this distribution is affected by the interaction and suffers a width broadening controlled by ${w}_{M}$. We also note here that there is no superkick-related effect and that the distribution is symmetrical for the inversion of the $y$ momentum.

To prove that the random momentum is indeed exchanged between the photon and the mask we can also compute the marginal distribution for the sum of the two momenta, that shows no broadening but only the expected combination of the two initial widths for the photon and mask momenta:
\begin{equation}
\begin{aligned}
f_T\left(\bm q_{\mathrm{tot}}=\bm q+\bm q_C\right)
&=\frac{4A_g^2w_0^4w_C^2w_M^2}
{\pi\left(w_0^2+w_C^2\right)^2
\left(w_0^2+w_C^2+w_M^2\right)\hbar^2}
e^{
-\frac{2d_M^2}{w_0^2+w_C^2+w_M^2}
-\frac{w_0^2w_C^2
\left(q_{\mathrm{tot},x}^2+q_{\mathrm{tot},y}^2\right)}
{2\left(w_0^2+w_C^2\right)\hbar^2}
}\\
&\quad\times
\left[
\frac{w_C^4
\left(q_{\mathrm{tot},x}^2+q_{\mathrm{tot},y}^2\right)}
{4\hbar^2\left(w_0^2+w_C^2\right)}
\pm\frac{d_Mw_C^2q_{\mathrm{tot},y}}
{\hbar\left(w_0^2+w_C^2+w_M^2\right)}
+\frac{
w_M^2\left(w_0^2+w_C^2+w_M^2\right)
+2d_M^2\left(w_0^2+w_C^2\right)}
{2\left(w_0^2+w_C^2+w_M^2\right)^2}
\right].
\end{aligned}
\end{equation}

\begin{figure}[th]
  \centering
  \includegraphics[width=\textwidth,keepaspectratio]
    {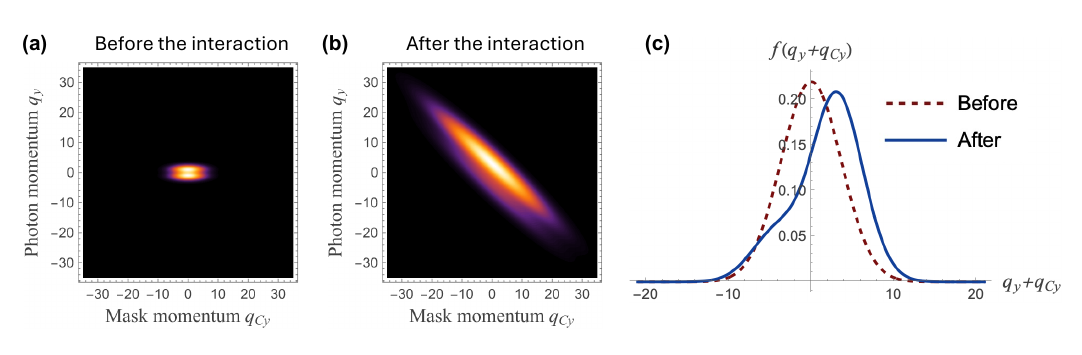}
    \caption{Analysis of the $q_y$ momentum balance in the mask-photon interaction, for $\frac{{w}_{M}}{{w}_{0}}=0.1, \frac{{d}_{M}}{{w}_{0}}=0.1,\frac{{w}_{C}}{{w}_{0}}=0.3$. All momentum values are given in units of $\hbar /{w}_{0}$. {\bf(a)-(b)} show two-dimensional plots of the joint marginal distribution for the ${q}_{y},{q}_{Cy}$ momenta before {\bf(a)} and after {\bf(b)} the interaction (conditional on photon transmission). {\bf(c)} shows the marginal distribution for the total momentum $q_{\mathrm{tot},y}=q_y+q_{Cy}$ before (dashed line) and after (solid line) the interaction (for the latter, the distribution is conditional on photon transmission).}
    \label{figS:mask-recoil}
\end{figure}

Figure \ref{figS:mask-recoil}{\bf(a)-(b)} shows two-dimensional plots of the joint marginal distribution for the ${q}_{y},{q}_{Cy}$ momenta before and after the interaction (conditional on photon transmission), to further illustrate this behavior. It is evident from these figures that the interaction between the photon and the mask induces broadening of the distributions for both the photon and the mask, but these two distributions become anti-correlated, because the random momentum is exchanged between them (this is the ordinary recoil effect). If we consider the total mask-photon momentum, this ``internal'' momentum exchange should be canceled out. Fig.~\ref{figS:mask-recoil}{\bf(c)} shows the $q_y$-marginal distribution for this total momentum before and after the interaction (conditional on photon transmission). We can see that in this plot there indeed is no random-broadening effect, but the ``superkick'' momentum shift appears, nevertheless. This is because this momentum is not coming from the photon-mask interaction.

The total momentum shown in Fig.~\ref{figS:mask-recoil}{\bf(c)} should also correspond to the momentum for an absorbing mask, as opposed to a transmitting mask, which in turn would correspond to the momentum distribution for an atom absorbing a photon from the superoscillating region of the wavefunction and receiving a superkick from it, as discussed by Berry and Barnett in their 2013 paper \cite{barnett_superweak_2013}.

\section{Experimental apparatus}
\label{app:supp-setup}

\subsection{Source}
Our photon source is a superluminescent diode (SLED) from Thorlabs (SLD810S) coupled to a single mode fiber. Its spectrum is inherently broadband ($\sim 25\,\mathrm{nm}$), so a $3\,\mathrm{nm}$ bandpass filter (Semrock LL01-810-25) is used to define a narrow bandwidth. Polarization is cleaned to match the optical axes of the spatial-light modulators using a polarising beam splitter, and the beam is magnified and collimated using a pair of lenses ($75\,\mathrm{mm}$--$200\,\mathrm{mm}$).

\subsection{Input state preparation}
The first spatial light modulator (Meadowlark 1920x1200 S-Series) is used to prepare the desired input beam following the method introduced by Bolduc et al.~\cite{bolduc_exact_2013}. This method is used to find the required hologram to generate the Fourier transform of the target field at the first diffraction order, given a blazed grating. This is done via a lens ($f=150\,\mathrm{mm}$) and an iris/slit, selecting only the first order. Using this method, we prepare an input LG beam with widths ranging in the $0.75$--$0.90\,\mathrm{mm}$ interval. To avoid unwanted diffraction effects from the edge of the SLM active area, we set the width of the incident beam, $w_{\mathrm{in}}$, to be $1.6\,\mathrm{mm}$. Since the masking method of \cite{bolduc_exact_2013} assumes a plane-wave illumination, we must compensate for this finite width. For a Gaussian or LG$_{01}$ beam with a target width $w_{\mathrm{tar}}$, we have
\begin{equation}
  w_{\mathrm{hologram}}
  = \frac{w_{\mathrm{tar}}w_{\mathrm{in}}}{\sqrt{w_{\mathrm{in}}^{2} - w_{\mathrm{tar}}^{2}}}.
\end{equation}

\begin{figure}[th]
  \centering
  \includegraphics[width=0.7\textwidth,keepaspectratio]
    {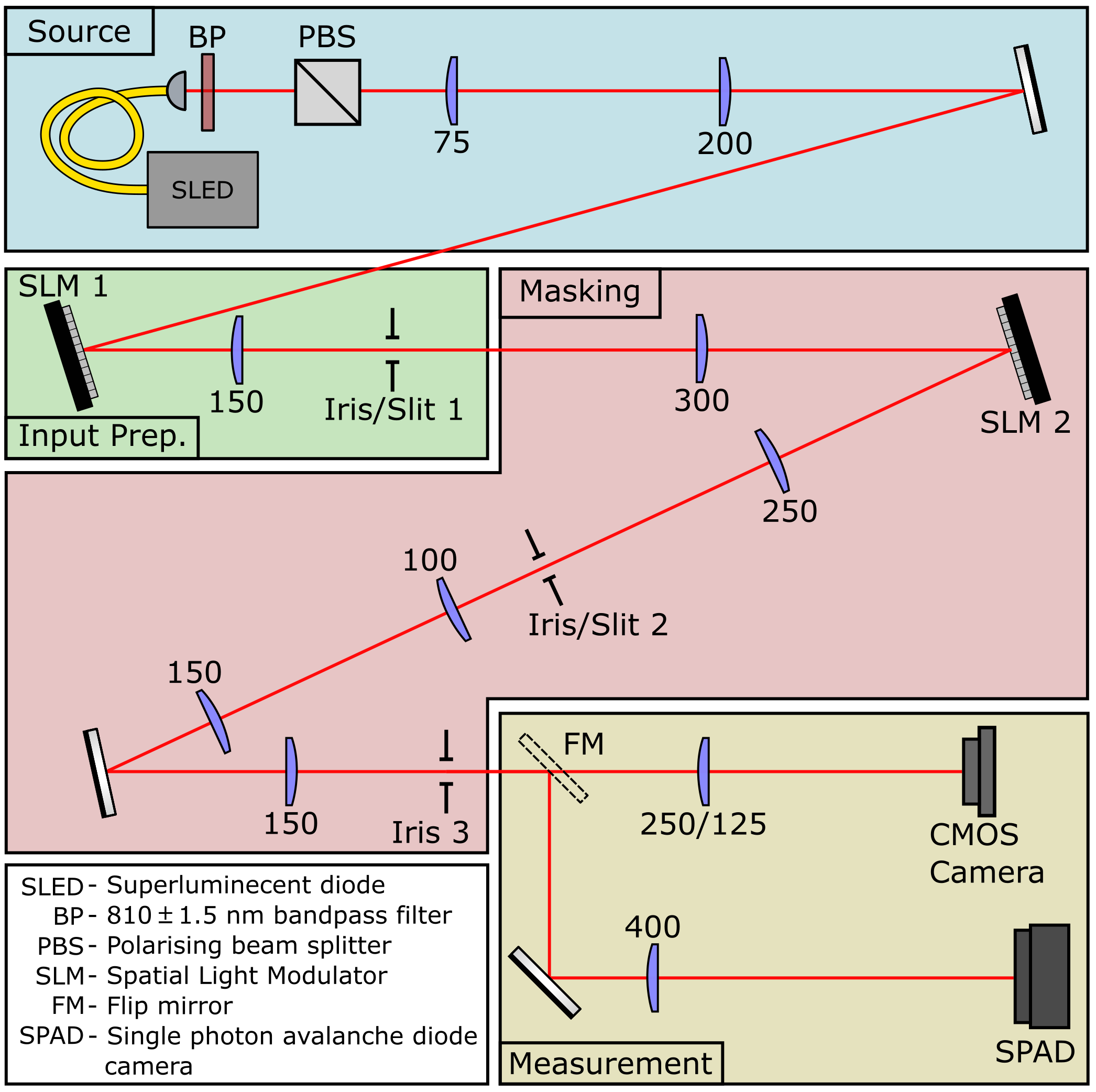}
    \caption{Schematic of the experimental apparatus. Lens focal lengths are in mm.}
    \label{figS:setup}
\end{figure}

\subsection{Masking}
The second SLM (also Meadowlark 1920x1200 S-Series) is used to apply the mask to the input state. Following the iris after SLM 1, another lens ($f=300\,\mathrm{mm}$) is used to Fourier-image the iris plane onto SLM 2, effectively magnifying the target field generated by SLM 1 by a factor of 2, giving $w_{0}$ ranging from $1.5$--$1.8\,\mathrm{mm}$. This magnification allows a small unclipped field at SLM 1 to fill the display of SLM 2 where pixelation can limit the mask size. The mask is realized as a contrast-modulated binary phase grating whose diffraction amplitude has a radial Gaussian profile with width $w_{M} \approx 0.1w_{0}$, centered at $x_{M}=\pm d_{M}$ with $d_{M}\approx 0.1w_{0}$, relative to the input beam center. A lens ($f=250\,\mathrm{mm}$) is used to separate the diffraction orders and a slit selects the first order, containing only the photons transmitted by the mask. The masked binary grating is defined explicitly as
\begin{equation}
  G_{\mathrm{mask}}(x,y) = H(x,y)G_{\mathrm{bin}}(x,y),
\end{equation}
with
\begin{equation}
  H(x,y) = \frac{2}{\pi}\sin^{-1}\!\left(A(x,y)\right),
\end{equation}
for a given \emph{amplitude} transmission profile $A(x,y)\in[0,1]$, and binary grating $G_{\mathrm{bin}}(x,y)\in\{0,\pi\}$ is any periodic binary function, for example $G_{\mathrm{bin}}(x,y)=\pi \times \mathrm{round}[0.5(1+\sin(2\pi y/\Lambda))]$.

Due to slight imperfections of the SLM, some additional scattered light is present, which results in some coherent speckle effects in the final far-field distributions. Another iris is placed in a near-field plane of SLM 2, centered on the mask position, which blocks the majority of this scattered light. The radius of the iris aperture is $> 5w_{M}$, and so has negligible effect on the action of the mask itself. To further reduce the impact of this scattered light, the final distributions are averaged over gratings with varying starting pixels, $G_{\mathrm{mask}}^{i}(x,y)$. Specifically,
\begin{equation}
  G_{\mathrm{mask}}^{i}(x,y)
  = H(x,y)G_{\mathrm{bin}}^{i}(x,y)
  = H(x,y)G_{\mathrm{bin}}(x,y-i)
\end{equation}
(for gratings oscillating in the $y$-direction).

\subsection{Detection}
The final momentum distributions were measured using cameras placed in the far-field of SLM 2, following suitable imaging and magnification such that the distributions covered a sufficiently large area of the sensor. A standard CMOS camera (Thorlabs Zelux CS165MU/M) was used for alignment and initial checks, while a single-photon avalanche diode (SPAD) array (PiImaging SPAD512) was used to measure the final momentum distributions in the single-photon regime. The SPAD camera was operated in a 1-bit detection mode such that, in each frame, a pixel returned 0 if no photon was detected, and 1 if a photon was detected.

SPAD arrays typically contain a fraction of dead or hot pixels that are effectively always (or the majority of the time) in the ``on'' state. They do not return any useful information, and so should be managed to avoid them introducing errors. These hot pixels are characterised by considering the photon-number statistics over a large number of frames, with the camera cap on so only dark counts should be detected. We sum this large number ($>10^{5}$) of frames to give a final image, or map of the dark counts measured in the acquisition time. We then define a threshold value, $N_{\mathrm{th}}$, and count the fraction of pixels with counts above this threshold,
\begin{equation}
  P_{\mathrm{meas}}^{\mathrm{th}}
  = P_{\mathrm{meas}}\!\left(N>N_{\mathrm{th}}\right).
\end{equation}
After removing these above-threshold pixels, we calculate the mean value of the remaining pixels, $\mu_{\mathrm{th}}$. We can then compute the fraction of above-threshold pixels expected assuming the true dark counts are Poisson-distributed with mean $\mu_{\mathrm{th}}$, denoted
\begin{equation}
  P_{\mathrm{theory}}^{\mathrm{th}}
  = P_{\mathrm{theory}}\!\left(N>N_{\mathrm{th}}\right).
\end{equation}
Since the hot pixels should skew the measured statistics, we can find the correct threshold to remove them by comparing $P_{\mathrm{meas}}^{\mathrm{th}}$ and $P_{\mathrm{theory}}^{\mathrm{th}}$. If $N_{\mathrm{th}}$ is set correctly, then we should find $P_{\mathrm{meas}}^{\mathrm{th}}\approx P_{\mathrm{theory}}^{\mathrm{th}}$. On the other hand, if $N_{\mathrm{th}}$ is too high, then some hot pixels will remain, which will skew $\mu_{\mathrm{th}}$ to be high, resulting in $P_{\mathrm{theory}}^{\mathrm{th}}>P_{\mathrm{meas}}^{\mathrm{th}}$. Conversely, if $N_{\mathrm{th}}$ is too low, then some true dark count values will be discounted, skewing $\mu_{\mathrm{th}}$ to be low, resulting in $P_{\mathrm{meas}}^{\mathrm{th}}>P_{\mathrm{theory}}^{\mathrm{th}}$. After finding the correct $N_{\mathrm{th}}$, we define the map of hot pixels, and either set these to 0 in subsequent measurements, or set them to be equal to the mean of their four direct neighbours.

\subsection{Single-photon regime}
To demonstrate the superkick effect in the regime of individual quantum events, measurements at the single-photon level are required. We use an attenuated coherent state to approximate a single-photon state. This is done by setting the SLED power and individual frame exposure time such that the average number of photons per pixel per frame in the brightest region of the beam (set for each measurement configuration separately) was approximately 0.02. In this regime, for a coherent state, the fraction of detection events containing $>1$ photon is less than 1\%. In practice, we set the frame exposure time to the minimum allowed value of $20\,\mathrm{ns}$ and changed the SLED power to give the desired mean.

\section{Data analysis procedure}
\label{app:supp-data-analysis}

Each SPAD image, representing a two-dimensional (2D) momentum distribution, is obtained by accumulating detections from a large number of photons. The total number of detected photons per image ranges from a few million to a few hundred million. In all SPAD images, dead and hot pixels (see Appendix~\ref{app:supp-setup}) are replaced with the mean of their four nearest neighbours. The dark-count background is estimated from a SPAD image acquired with the light blocked.

Each complete dataset used to test local momentum conservation comprises four distinct two-dimensional momentum-distribution images: MD1, the unmasked input LG beam; MD2, the mask-transmitted LG beam; MD3, an unmasked Gaussian beam; and MD4, the mask-transmitted Gaussian beam. MD1 and MD2 provide the primary data: MD1 gives the input momentum distribution, whereas MD2 gives the momentum distribution of the photons transmitted through the mask. MD3 and MD4 are used together to determine the actual mask width $w_M$ from the change in the momentum-distribution width. MD4 also provides the experimental mask-kick distribution $K_M(\bm{q})$. To construct the experimental MLMC reference distribution, we extract the selected marginal distribution $f_S(q_y)$ from MD1 and numerically convolve it with the corresponding marginal mask-kick distribution $K_M(q_y)$ obtained from MD4.

Let us consider MD1 in more detail. It records the two-dimensional momentum distribution of the unmasked input LG beam. To acquire this image, SLM2 displays a uniform grating that directs the beam toward the detection line without applying any additional spatial modulation. The center and width of the far-field image are determined by fitting a theoretical LG intensity distribution to the data. The fitted center defines the origin of the $\bm{q}$ momentum plane and is kept fixed for all subsequent measurements. The fitted width is used to calibrate the momentum-per-pixel scale in units of $\hbar/w_0$, which is likewise kept fixed. The precise calibration depends on the imaging geometry. For the data presented here and in the main text, one unit of $\hbar/w_0$ corresponds to 5.38 pixels. Thus, the momentum increment between adjacent pixels is $\Delta q=(1/5.38)\hbar/w_0$. After background subtraction, each two-dimensional distribution is normalized by its measured total photon count and converted into a probability density per unit momentum area by dividing by $\Delta q^2$.

\begin{figure}[h]
  \centering
  \includegraphics[width=\textwidth,keepaspectratio]
    {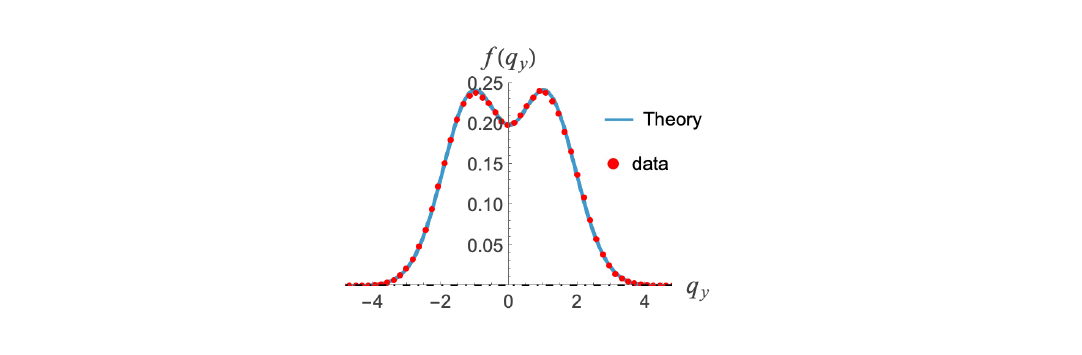}
    \caption{Marginal $q_y$ momentum distribution for the input LG mode.}
    \label{figS:LGprofile}
\end{figure}

\begin{figure}[h]
  \centering
  \includegraphics[width=\textwidth,keepaspectratio]
    {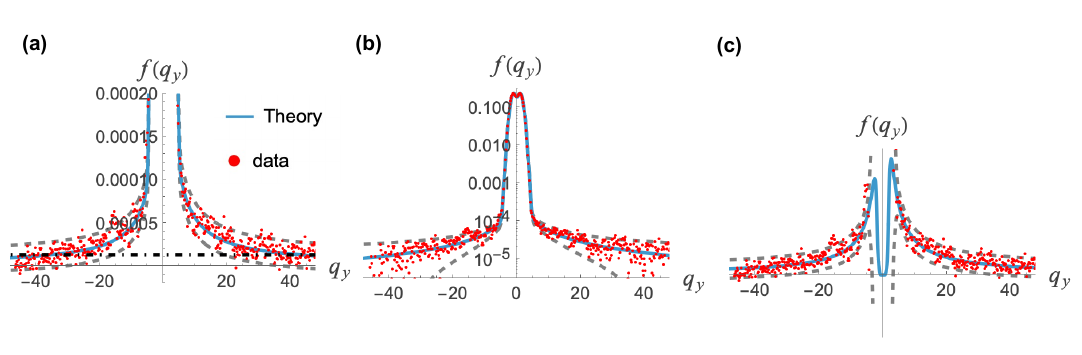}
    \caption{Pedestal appearing with the input LG mode. {\bf(a)} is the LG $q_y$ marginal zoomed to a very small scale. {\bf(b)} is the semilog plot of the $q_y$-marginal distribution. {\bf(c)} is a view of the same data for $|q_y|>6$ together with the theoretical curve given by Eq.~\protect\eqref{eq:pedestal}. The two dashed grey lines give a two-sigma variation, based on Poissonian statistics only, from theory. The dot-dashed line in {\bf(a)} is the two-sigma background noise level.}
    \label{figS:pedestal}
\end{figure}
Examples of the input LG momentum distribution are shown in Fig.~\ref{fig:inputLG} of the main text, and its $q_y$ marginal is shown in Fig.~\ref{figS:LGprofile}. This marginal distribution, like those presented below, is obtained by summing the two-dimensional probability density over the $q_x$ direction and multiplying by the momentum increment $\Delta q$. At the vertical scale used in Fig.~\ref{figS:LGprofile}, small imperfections are not visible, and the agreement between data and theory appears nearly perfect. However, magnifying the vertical scale of the $q_y$ marginals by a factor of $10^3$ or using a logarithmic scale reveals a weak pedestal superimposed on the LG distribution (see Fig.~\ref{figS:pedestal}). We attribute this pedestal to apodization by finite-aperture optics in intermediate propagation planes, possibly combined with a weak coherent optical background. At large $|q_y|$, its asymptotic behavior is well fitted by a $1/|q_y|$ dependence (apodization alone, without interference with a background, is expected to produce a $1/|q_y|^2$ dependence). We fit the pedestal over the full range of $q_y$ using the following phenomenological expression:
\begin{equation}
f_p(q_y)=\left[A_{pR}\theta(q_y)+A_{pL}\theta(-q_y)\right] \frac{q_p q_y^n}{q_p^{n+1} + |q_y|^{n+1}}+B_p.
\label{eq:pedestal}
\end{equation}
where $n\ge2$ is an even integer. This functional form is chosen so that its nonconstant terms reproduce the required asymptotic $1/|q_y|$ behavior for $|q_y|\gg q_p$ while vanishing smoothly near the origin to avoid affecting the maximum LG values, as shown in Fig.~\ref{figS:pedestal}{\bf(c)}. The adjustable amplitudes $A_{pL}$ and $A_{pR}$ are allowed to differ on the negative- and positive-$q_y$ sides, respectively. We also include a constant background $B_p$ as a fit parameter to account more accurately for background variations caused by stray light from the SLMs and other optical components. After some testing, the parameters $n$ and $q_p$ in Eq.~\eqref{eq:pedestal} have been fixed as $n=6$, for having a sufficiently rapid vanishing of the pedestal model in the central region, and $q_p=2\hbar/w_0$, to match smoothly the tails of the LG mode. The fit is restricted to data with $|q_y|>6\hbar/w_0$, where the contribution of the LG mode is negligible. To further reduce possible bias, we subtract the theoretical LG marginal from the data before fitting. Examples of the resulting best fits are shown in Fig.~\ref{figS:pedestal}. The pedestal fit parameters obtained for the data shown in the main article and in Fig.~\ref{figS:pedestal} are reported in Table \ref{tab:pedestal-fit-parameters}.

\begin{table}[h]
    \centering
    \caption{Pedestal fit parameters obtained for the data shown in the main article (after correcting for the actual background based on a previous run of the same fit), together with their standard deviations determined from the fit.}
    \label{tab:pedestal-fit-parameters}
    %\small
    %\footnotesize
    \begin{tabular}{cc}
        \hline
        Fit parameter & best-fit value \\
        \hline
        $A_{pL}$ & $(2.32\pm0.09)\times10^{-4} w_0/\hbar$ \\
        $A_{pR}$ & $(2.80\pm0.09)\times10^{-4} w_0/\hbar$ \\
        $B_{p}$ & $(6\pm8)\times10^{-7} w_0/\hbar$ \\
        \hline
    \end{tabular}
\end{table}

\begin{figure}[h]
  \centering
  \includegraphics[width=\textwidth,keepaspectratio]
    {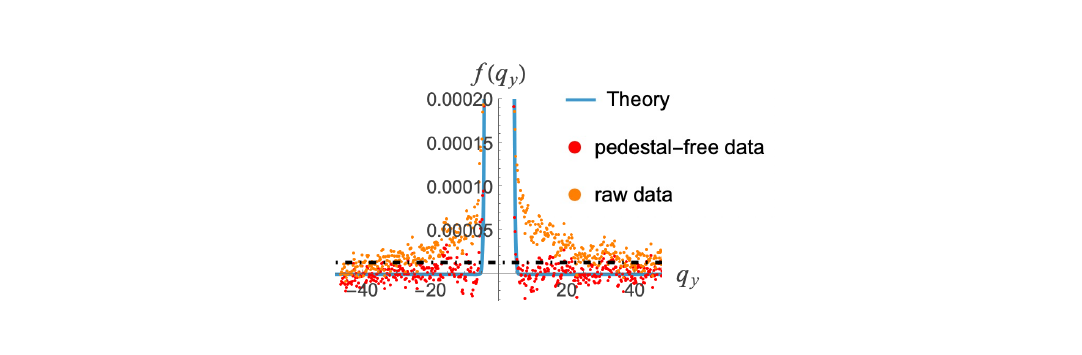}
    \caption{Marginal $q_y$ momentum distribution of the input LG mode before (orange dots) and after (red dots) subtraction of the fitted pedestal. The theory line in this case is the pure LG mode. The dot-dashed line is the two-sigma background noise level, based on Poissonian statistics.}
    \label{figS:pedestalfree}
\end{figure}
Once the theoretical pedestal curve $f_p(q_y)$ has been determined, we subtract it from the data to obtain the ``pedestal-free'' distribution shown in Fig.~\ref{figS:pedestalfree}. This subtraction is necessary because the pedestal would otherwise strongly distort the statistical properties of the input distribution, particularly the selected tails $f_S$ used to reconstruct the MLMC reference distribution. This remains true even when the pedestal amplitude is extremely small, because its asymptotic $1/|q_y|$ behavior is nonintegrable. Including the pedestal would therefore make derived statistical quantities, such as the mean momentum or cumulative probabilities, depend on the arbitrarily chosen experimental window.

\begin{figure}[h]
  \centering
  \includegraphics[width=\textwidth,keepaspectratio]
    {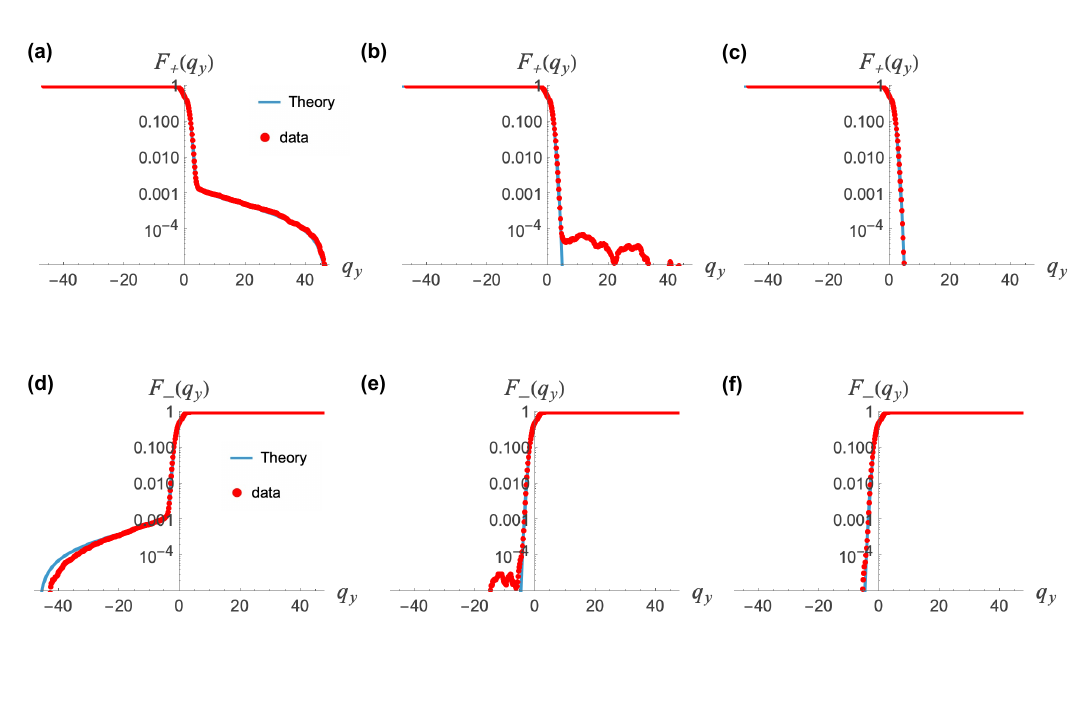}
    \caption{Cumulative marginal momentum distributions of the input LG mode under different pedestal treatments. Panels {\bf(a)--(c)} show the upper cumulative distribution $F_+(q_y)$, whereas panels {\bf(d)--(f)} show the lower cumulative distribution $F_-(q_y)$. Panels {\bf(a)} and {\bf(d)} use the original data; panels {\bf(b)} and {\bf(e)} show the data after subtraction of the fitted pedestal; and panels {\bf(c)} and {\bf(f)} show the data after the additional removal of residual fluctuations.}
    \label{figS:cumulative}
\end{figure}
Next, we define the experimental lower and upper cumulative distributions as $F_-(q_y)=\int_{-\infty}^{q_y}f(q_y')\,dq_y'$ and $F_+(q_y)=1-F_-(q_y)=\int_{q_y}^{+\infty}f(q_y')\,dq_y'$, respectively. These functions are used to determine the threshold $q_0$ that defines the selected tail distribution $f_S(q_y)$ employed in the construction of the MLMC reference. Figure~\ref{figS:cumulative} compares the cumulative distributions obtained from the original data, the pedestal-subtracted data, and the pedestal-subtracted data after additionally removing the residual background fluctuations. The latter operation is done by removing all data points lying below three standard deviations of the background Poissonian noise and possible data points fluctuating back above this noise threshold far from the main signal, i.e. for $|q_y|>9\hbar/w_0$ or higher; possible data fluctuations that remained adjacent to the main LG signal were not removed by this procedure and contributed to the LG tail imperfections discussed below.

\begin{figure}[h]
  \centering
  \includegraphics[width=\textwidth,keepaspectratio]
    {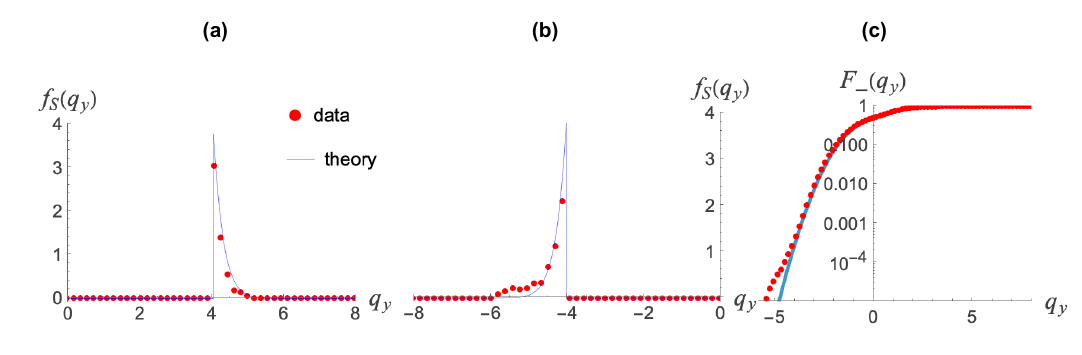}
    \caption{Selected momentum distributions constructed from the tails of the input LG mode. {\bf(a)} Distribution selected from the positive-$q_y$ tail. {\bf(b)} Distribution selected from the negative-$q_y$ tail, retaining a deviation from the ideal LG profile. {\bf(c)} Enlarged view of the corresponding feature in the cumulative distribution.}
    \label{figS:selectedf}
\end{figure}
The resulting experimental selected distributions are shown in Fig.~\ref{figS:selectedf}. The positive-$q_y$ tail agrees closely with the ideal LG distribution for $q_y>q_0$. By contrast, the negative-$q_y$ tail exhibits a clear deviation from the ideal profile for $q_y<-q_0$. Because this measured imperfection is retained when constructing the experimental MLMC reference, it shifts the reference toward more negative values of $q_y$, thereby reducing the discrepancy with the transmitted-photon distribution and making the test of local momentum conservation more conservative.

To assess the impact of the pedestal subtraction procedure on all final estimated uncertainties, we propagate pedestal-fit uncertainty with 1000 Monte Carlo repetitions. In each repetition, the pedestal parameters are drawn from their estimated joint distribution and the complete subtraction and MLMC reconstruction are repeated. The resulting spread contributes chiefly to the uncertainty in $\bar{q}_{Sy}$.

\begin{figure}[h]
  \centering
  \includegraphics[width=\textwidth,keepaspectratio]
    {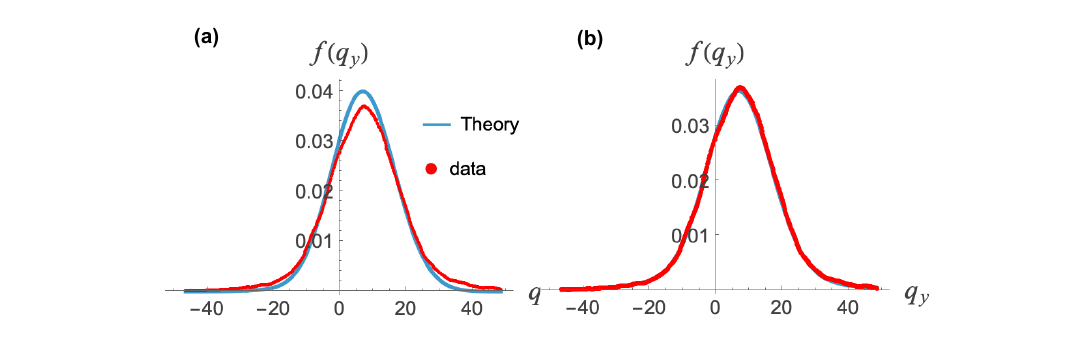}
    \caption{Marginal $q_y$ momentum distributions of mask-transmitted photons. {\bf(a)} Experimental data and the parameter-free theoretical prediction. {\bf(b)} The same data compared with the model including optical apodization, as defined by Eq.~\protect\eqref{eq:apodization}, which provides improved quantitative agreement. The best-fit apodization parameters are $\eta=0.20$ and $\gamma=12 \hbar/w_0$.}
    \label{figS:mask-transmitted}
\end{figure}
The $q_y$ marginal distribution obtained from image MD2 is shown in Fig.~\ref{figS:mask-transmitted}. Panel {\bf(a)} compares the data with the theoretical prediction, which contains no adjustable parameters. Although the overall agreement is very good, the measured peak is slightly lower and the tails are slightly higher than predicted. We attribute this discrepancy to optical apodization caused by finite apertures in the imaging optics of the detection line. This effect can be stronger than the apodization responsible for the pedestal in the input LG distribution because the masked beams are approximately ten times wider than the input beam (but remaining well within the paraxial regime). We model the effect of apodization on a distribution $f(q_y)$ using the following phenomenological expression:
\begin{equation}
f_{A}(q_y)=(1-\eta)f(q_y)+\eta K_A(q_y) \ast f(q_y),
\label{eq:apodization}
\end{equation}
where
\begin{equation}
K_A(q_y)=\frac{\gamma}{\pi(\gamma^2+q_y^2)}.
\end{equation}
Here, $\ast$ denotes convolution, while $\eta$ and $\gamma$ are fit parameters. The resulting best fit is shown in Fig.~\ref{figS:mask-transmitted}{\bf(b)}.

\begin{figure}[h]
  \centering
  \includegraphics[width=\textwidth,keepaspectratio]
    {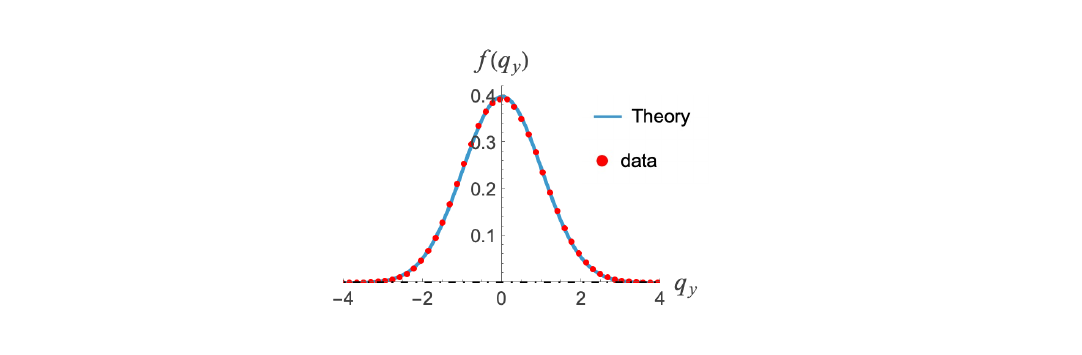}
    \caption{Marginal $q_y$ momentum distribution of the unmasked Gaussian beam, compared with the theoretical prediction.}
    \label{figS:gaussian}
\end{figure}
Image MD3 records the momentum distribution of an unmasked Gaussian beam with nominal waist $w_0$, corresponding to a momentum-space width $w_q=2\hbar/w_0$. As for MD1, during this measurement, SLM2 displays a uniform grating and just directs the beam towards detection. MD3 is used together with MD4 to determine the mask width precisely. Figure~\ref{figS:gaussian} shows the measured $q_y$ marginal distribution and the corresponding theoretical prediction.

\begin{figure}[h]
  \centering
  \includegraphics[width=\textwidth,keepaspectratio]
    {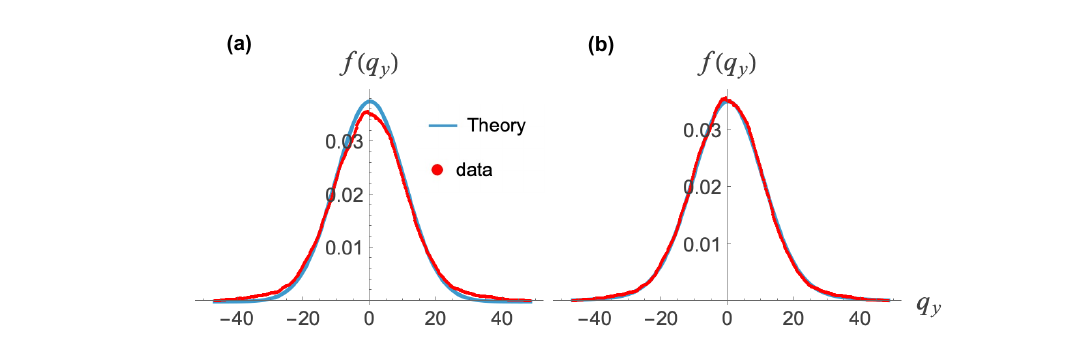}
    \caption{Marginal $q_y$ momentum distributions of the Gaussian beam transmitted through the mask. {\bf(a)} Experimental data and the parameter-free theoretical prediction. {\bf(b)} The same data compared with the model including optical apodization, as defined by Eq.~\protect\eqref{eq:apodization}, which provides improved quantitative agreement. The best-fit apodization parameters are $\eta=0.14$ and $\gamma=14\hbar/w_0$.}
    \label{figS:maskedGaussian}
\end{figure}
Image MD4 records the momentum distribution of the Gaussian beam transmitted through the mask. Its $q_y$ marginal is shown in Fig.~\ref{figS:maskedGaussian}. We fit the two-dimensional MD4 distribution with a Gaussian whose width is an adjustable parameter. Comparing the fitted width with the theoretical value $\sqrt{w_q^2+w_{Mq}^2}$, where $w_{Mq}=2\hbar/w_M$, allows us to determine the actual mask width $w_M$ with a relative precision of $10^{-2}$ or better. After normalization, MD4 also provides the experimental mask-kick distribution $K_M$. Strictly, $K_M$ should be measured using a plane-wave input; however, the correction associated with using a Gaussian input is negligible and slightly broadens the inferred $K_M$, making the test of local momentum conservation more conservative. Including optical apodization further improves the agreement between data and theory, as shown in Fig.~\ref{figS:maskedGaussian}{\bf(b)}.

\begin{figure}[h]
  \centering
  \includegraphics[width=\textwidth,keepaspectratio]
    {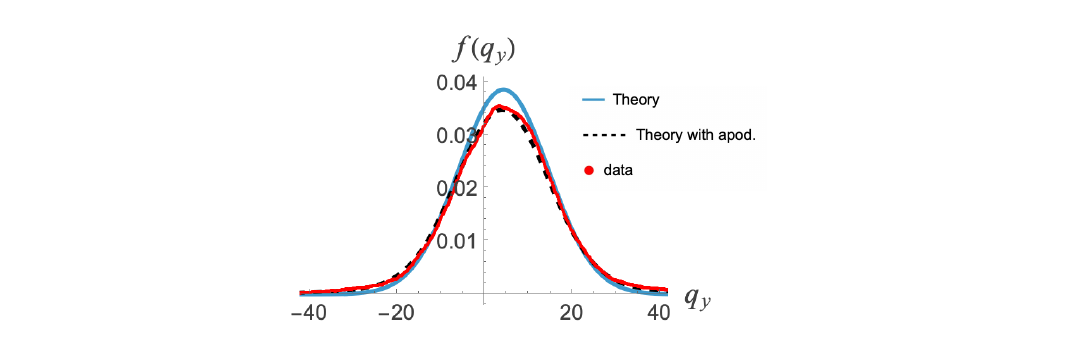}
    \caption{Experimentally reconstructed MLMC distribution compared with the theoretical predictions without and with optical apodization. The latter retains the same apodization model already considered for MD4, without modification of the best-fit parameters.}
    \label{figS:expMLMC}
\end{figure}

Finally, we construct the experimental MLMC reference distribution by numerically convolving the experimental selected distribution $f_S$ with the experimental mask-kick distribution $K_M$. Figure~\ref{figS:expMLMC} compares the resulting distribution with the theoretical MLMC predictions obtained without and with optical apodization.

\end{document}